\documentclass[twocolumn,showpacs,preprintnumbers,amsmath,amssymb]{revtex4}
\usepackage{graphicx}% Include figure files
\usepackage{dcolumn}% Align table columns on decimal point
\usepackage{bm}% bold math

\def\bea{\begin{eqnarray}}
\def\eea{\end{eqnarray}}

\def\pp{\mbox{$p$-$p$} }
\def\auau{\mbox{Au-Au} }

\def\aa{\mbox{A-A} }
\def\nn{\mbox{N-N} }

\def\ee{\mbox{$e^+$-$e^-$} }

\def\deta{$\eta_\Delta$ }
\def\dphi{$\phi_\Delta$ }
\def\drho{$\Delta \rho / \sqrt{\rho_{ref}}$ }
\def\ytyt{$y_t \times y_t$ }
\def\ptpt{$p_t \times p_t$ }

\begin{document} 

\preprint{Version 1.5}

\title{Gluon correlations from a Glasma flux-tube model compared to measured hadron correlations on transverse momentum $\bf (p_t,p_t)$ and angular differences $\bf (\eta_\Delta,\phi_\Delta)$
}

\author{Thomas\ A.\ Trainor}
\address{CENPA 354290, University of Washington, Seattle, Washington 98195}
\author{R.\ L.\ Ray}
\address{Department of Physics, University of Texas, Austin, Texas  78712}

%%%%%%%%%%%%%%%%%%%%%%%%%%%%%%%

\date{\today}

\begin{abstract}
A Glasma flux-tube model has been proposed to explain strong elongation on pseudorapidity $\eta$ of the same-side 2D peak in minimum-bias angular correlations from $\sqrt{s_{NN}} = 200$ GeV \auau collisions. The same-side peak or ``soft ridge'' is said to arise from coupling of flux tubes to radial flow. Gluons radiated transversely from flux tubes are boosted by radial flow to form a narrow structure or ridge on azimuth. In this study we test the conjecture by comparing predictions for particle production, spectra and correlations from the Glasma model and conventional fragmentation processes with measurements. We conclude that the Glasma model is contradicted by measured hadron yields, spectra and correlations, whereas a two-component model of hadron production, including minimum-bias parton fragmentation, provides a quantitative description of most data, although $\eta$ elongation remains unexplained.
\end{abstract}


\maketitle

%%%%%%%%%
 \section{Introduction}

The systematics of measured hadron production and multihadron correlations in \pp and more-peripheral \aa collisions are described quantitatively by a two-component model including soft and hard components~\cite{ppprd,porter1}. By hypothesis the soft component arises from longitudinal fragmentation of participant projectile nucleons as a result of soft momentum transfers (e.g.\ diffractive scattering). The hard component arises from minimum-bias large-angle parton scattering and transverse fragmentation as a result of (semi)hard momentum transfers. Those interpretations have been tested and elaborated in several studies~\cite{porter2,porter3,axialci,daugherity,hardspec,fragevo,jetspec}.

Conventional theoretical descriptions of soft and hard components combine parton distribution functions (PDFs) [hard component, perturbative QCD (pQCD)] with ``limiting fragmentation'' (parton splitting cascade, DGLAP~\cite{dglap1,dglap2}) at larger momentum fraction $x$ and a phenomenological nonperturbative approach (e.g.\ Lund string model) at smaller $x$~\cite{lund}. Examples of Monte Carlo models combining both aspects are PYTHIA for \pp collisions~\cite{pythia} and HIJING for \aa collisions~\cite{hijing}.

An alternative description of longitudinal particle production in more-central \aa collisions is based on the  Color Glass Condensate (CGC)~\cite{cgc}. The CGC model invokes a statistical ensemble of classical color charges (sources) at larger $x$ and a radiated classical color field at smaller $x$ described as a {\em Glasma}. There is obviously a correspondence between Glasma vs Lund strings at smaller $x$, and color radiators vs pQCD parton splitting cascade at larger $x$. We wish to explore those dichotomies by quantitative comparisons of theory with spectrum and two-hadron correlation data. 

In a previous study we considered the relation between Glasma flux-tube predictions and angular-correlation phenomenology for same-side (defined below) correlations in \pp and \auau collisions, specifically the so-called ``soft ridge''~\cite{tomglasma}. The study concluded that whereas pQCD-based descriptions of jet-related angular correlations are in quantitative agreement with data the description based on Glasma flux tubes is inconsistent with correlation data in several ways.

In the present study we consider the relation between Glasma predictions~\cite{lappi,lappi2} and measured two-particle (gluon and hadron) correlations on transverse momentum $p_t$ (or transverse rapidity $y_t$ defined below) as well as pseudorapidity $\eta$ and azimuth $\phi$. We find further substantial discrepancies between CGC-Glasma predictions and data.
We conclude that even if conjectured radial flow played a role in nuclear collisions the Glasma flux-tube model does not correspond to physical processes that might, in the presence of such flow, produce the observed same-side 2D peak structure elongated on $\eta$.

This article is arranged as follows: Sec.~\ref{analysis} reviews analysis methods applied to RHIC hadron data, Sec.~\ref{gluecorr} describes gluon correlations predicted by the Glasma flux-tube model, Sec.~\ref{hadcor} describes measured hadron correlation and spectrum data, and Sec.~\ref{glasmavs} compares predicted Glasma gluon correlations with the hadron data.

%%%%%%%%%
 \section{Analysis method} \label{analysis}

We review technical aspects of STAR correlation analysis applied to nuclear collisions at the Relativistic Heavy Ion Collider (RHIC). Method details are provided in Refs.~\cite{porter2,porter3,inverse,axialci,daugherity,davidhq,davidhq2,davidaustin}. Topics include 
\aa collision geometry, 
the two-component model of hadron production, 
correlation measures, 
2D histograms, 
model functions for 2D $\chi^2$ fits and the relation between fluctuations and angular correlations.

\subsection{\aa collision geometry} \label{ppcent}

\aa collision geometry is described by the Glauber model relating the \aa differential cross section to participant nucleon number $N_{part}$ and \nn binary-collision number $N_{bin}$~\cite{powerlaw}. A derived projectile-nucleon mean path length $\nu = 2 N_{bin} / N_{part}$ is also defined. Through the measured \aa differential cross section on charged-hadron multiplicity $n_{ch}$ within some angular acceptance the Glauber  parameters can be related to observed $n_{ch}$.

%\subsection{The Glauber model and low-$\bf x$ glue structure}

Optical $\epsilon_{opt}$~\cite{davidhq} and Monte Carlo $\epsilon_{MC}$~\cite{phobosquad} eccentricities have been invoked  to model \aa eccentricity required for interpretation of the azimuth quadrupole measured by $v_2$. 
The former assumes a smooth matter distribution across nuclei whereas the latter assumes that point-like participant nucleons are the determining elements.
{\em A priori} support for $\epsilon_{opt}$ assumes that the azimuth quadrupole emerges from interactions at small $x < 0.01$ where one might expect onset of a smooth, saturated glue system (e.g.\ Glasma)~\cite{gluequad}.
%. Large-$x$ nucleon structure should then be unimportan 
{\em A posteriori} support for $\epsilon_{opt}$ arises from a simple systematic trend $ \propto N_{bin}\epsilon^2_{opt}$ observed for $v^2_2\{2D\}$ data which accurately exclude contributions from jet structure (nonflow)~\cite{davidhq}. 
%The simple data trend extends to \pp collisions.

%, introducing a large sampling noise contribution $\sim 1/N_{part}$ to $\epsilon^2_{MC}$, especially notable in more-peripheral and more-central \aa collisions.

\subsection{Two-component hadron production model}

%CHECK REDUNDANCY WITH INTRODUCTION

According to the two-component model spectra and correlations from nuclear collisions can be decomposed (near mid-rapidity) into soft and hard components denoting respectively longitudinal fragmentation (mainly diffractive dissociation) of projectile nucleons and transverse fragmentation of large-angle-scattered partons~\cite{ppprd,hardspec,porter2,porter3}. Soft and hard components from \pp collisions are clearly distinguishable in \ptpt or \ytyt correlations (defined below). 

In more-peripheral \aa collisions the soft component should vary with centrality $\propto N_{part}$ and the hard component should vary $\propto N_{bin}$~\cite{kn}, those trends constituting the Glauber linear superposition reference for spectra and correlations. The soft-component correlation amplitude drops to zero in high-multiplicity \pp and more-central \auau collisions. The hard-component fraction is a few percent in minimum-bias (NSD) \pp collisions~\cite{ppprd} but increases to about one third of the final-state hadron yield in central \auau collisions~\cite{jetspec}. Thus, the two-component model of hadron spectra and correlations provides a self-consistent quantitative description of almost all aspects of nuclear collisions based on pQCD and measured properties of elementary \ee and \pp collisions.

\subsection{Correlation measures} \label{corrmeas}

Two-particle correlations are structures in pair-density distributions on six-dimensional momentum space $(p_{t1},\eta_1,\phi_1,p_{t2},\eta_2,\phi_2)$. We visualize correlation structure in 2D subspaces $(p_t,p_t)$ and $(\eta_\Delta,\phi_\Delta)$ (defined below) which retain almost all structure within a limited $\eta$ acceptance such as the STAR Time Projection Chamber (TPC)~\cite{tpc}.
We measure correlations with {\em per-particle} statistic  $\Delta \rho / \sqrt{\rho_{ref}} = \rho_0\, (\langle \hat r \rangle-1)$, where $\Delta \rho= \rho - \rho_{ref}$ is the correlated-pair density, $\rho_{ref}$ is the reference- or mixed-pair density, $\langle \hat r \rangle$ is the (unit-normal) sibling/mixed pair number ratio and prefactor  $\rho_0 = \bar n_{ch} / \Delta \eta\, \Delta \phi  $ is the charged-particle 2D angular density averaged over angular acceptance $(\Delta \eta,\Delta \phi)$~\cite{axialci,axialcd}. Pair ratio $\hat r$ is averaged over kinematic bins (e.g. multiplicity, $p_t$, vertex position), and we assume factorization of the reference pair density $\rho_{ref} \approx \rho_0^2$. 

The {per-particle} measure is an improvement over conventional {\em per-pair} correlation function $\langle \hat r \rangle \rightarrow \rho / \rho_{ref}$ or $ \langle \hat r \rangle - 1 \rightarrow \Delta \rho / \rho_{ref}$
since it eliminates a trivial $1/ n_{ch}$ trend common to all per-pair measures (except for quantum correlations). The intensive definition in terms of hadron 2D density $\rho_0(b)$ ($b$ is the \aa impact parameter) rather than multiplicity $n_{ch}$ also eliminates a trivial dependence on detector angular-acceptance factor $\Delta \eta \Delta \phi$.~\cite{inverse}. 
%
%The extensive form $n_{ch} (\langle \hat r \rangle - 1)$ includes a trivial acceptance factor $\Delta \eta \Delta \phi$ which depends on the specific detector configuration.

%Such measures have the property that under combination of two uncorrelated parts the per-particle measure value for the whole remains the same as for the parts. That property is desirable for tesing a Glauber \nn linear superposition (GLS) hypothesis for \aa centrality dependence.

%However, that {\em extensive} measure is proportional to the specific angular-acceptance product $\Delta \eta \Delta \phi$ defined for the analysis. 

%$\Delta \rho / \sqrt{\rho_{ref}}$  is invariant under combination of uncorrelated parts. The correlation measure is independent of angular acceptance if the underlying physical mechanisms are uniform across the acceptance. %Within the STAR TPC that is a good approximation~\cite{axialcd}. Within the larger CMS acceptance there are deviations from uniformity.

%For our initial angular correlation analysis we adopted $\langle  n_{ch} \rangle\, (\langle \hat r \rangle - 1)$ as the correlation measure, where $n_{ch}$ is the charged-particle multiplicity within the detector angular acceptance $(\Delta \eta,\Delta \phi)$~\cite{axialci,axialcd}. 

\subsection{Transverse-momentum correlations on $\bf p_t \times p_t$}

2D correlations on $p_t$ or transverse rapidity $y_t = \ln[(p_t + m_t) / m_\pi]$ ($m_\pi$ is assumed for unidentified hadrons) are complementary to 4D angular correlations in 6D two-particle momentum space. $y_t$ is preferred for visualizing correlation structure on transverse momentum. \ptpt or \ytyt and angular correlations can be defined for like-sign (LS) and unlike-sign (US) charge combinations and also for same-side (SS) and away-side (AS) azimuth subregions of angular correlations (defined below). Manifestations of different correlation mechanisms (e.g.\ soft and hard components) can be clearly distinguished in the four combinations of charge-pair type and azimuth subspace, with distinctive forms for each of the LS and US charge combinations and for SS and AS azimuth subspaces~\cite{porter2,porter3}.

For correlations on \ptpt the pair ratio is $\hat r(p_{t1},p_{t2})$ and the prefactor becomes $\sqrt{\rho_{ref}} = \sqrt{\rho_0(p_{t1})\, \rho_0(p_{t2})}$, the geometric mean of single-particle $p_t$ spectra. The correct prefactor is essential for proper comparison of predicted gluon correlations and measured hadron correlations.

\subsection{Angular correlations on $\bf (\eta_\Delta,\phi_\Delta)$}

Angular correlations can be formed by integrating over the entire \ptpt pair acceptance (minimum-bias angular correlations) or over subregions~\cite{porter2,porter3}. Examples of the latter include  ``trigger-associated'' dihadron correlations resulting from asymmetric cuts on \ptpt\cite{ptcuts}.

Two-particle angular correlations are defined on 4D momentum subspace $(\eta_1,\eta_2,\phi_1,\phi_2)$. Within acceptance intervals where correlation structure is invariant on mean angle (e.g.\ $\eta_\Sigma = \eta_1 + \eta_2$) angular correlations can be {\em projected by averaging} onto difference variables (e.g.\ $\eta_\Delta = \eta_1 - \eta_2$) without loss of information to form {\em angular autocorrelations}~\cite{axialcd,inverse}. The 2D subspace ($\eta_\Delta,\phi_\Delta$) is then visualized. 
%The notation $x_\Delta$ rather than $\Delta x$ for difference variables is adopted to conform to mathematical notation conventions and to reserve 
Symbol $\Delta x$ is used as a measure of the detector acceptance on parameter $x$. 

Angular correlations can be formed separately for like-sign  and unlike-sign charge combinations, as well as for the charge-independent (CI = LS + US) combination~\cite{axialci,axialcd}. The pair angular acceptance on azimuth can be separated into a same-side (SS) region ($|\phi_\Delta| < \pi / 2$) and an away-side (AS) region ($|\phi_\Delta| > \pi / 2$). The SS region includes {\em intra}\,jet correlations (hadron pairs within single jets), while the AS region includes {\em inter}\,jet correlations (hadron pairs from back-to-back jet pairs).

\subsection{Angular-correlation model functions}

The hard component of angular correlations includes a SS 2D peak at the angular origin and an AS 1D peak on azimuth uniform on $\eta_\Delta$ (within the STAR TPC acceptance). The minimum-bias SS 2D peak (intrajet correlations) is well modeled by a 2D Gaussian. 
Except for \pp and more-peripheral \aa collisions the AS peak (interjet correlations) is conveniently modeled as an AS dipole \mbox{$\propto \cos(\phi_\Delta - \pi)$}.
The soft component is modeled by a 1D Gaussian on $\eta_\Delta$ with r.m.s.\ width $\approx 1$. 

The combined model function including azimuth quadrupole term $\cos(2\phi_\Delta)$ required to describe \aa angular correlations within the STAR TPC  is~\cite{axialci,daugherity,davidhq}
\bea \label{modeleq}
\frac{\Delta \rho}{\sqrt{\rho_{ref}}}   & = &  A_0 
+ A_{\rm 1D}\, e^{-\frac{1}{2} \left( \frac{\eta_{\Delta}}{ \sigma_{\rm 1D}} \right)^2  } \\ \nonumber
&+& A_{\rm 2D} \, e^{- \frac{1}{2} \left\{ \left( \frac{\phi_{\Delta}}{ \sigma_{\phi_{\Delta}}} \right)^2    +    \left( \frac{\eta_{\Delta}}{ \sigma_{\eta_{\Delta}}} \right)^2 \right\}} \\ \nonumber
&+&   
A_{\rm D}\, \left[1+\cos(\phi_\Delta - \pi)\right]/2 + A_{\rm Q}\, 2\cos(2\, \phi_\Delta),
 %\nonumber \\
%&+& A_2 \, e^{- \left\{ \left( \frac{\phi_{\Delta}}{w_{\phi_{\Delta}}} \right)^2  + \left(\frac{\eta_{\Delta}}{ w_{\eta_{\Delta}}} \right)^2 \right\}^{1/2} }.
\eea
where a narrow 2D exponential describing quantum correlations and electron pairs from $\gamma$ conversions~\cite{daugherity} has been omitted for clarity.  The {\em nonjet} quadrupole amplitude is expressed in terms of conventional parameter $v_2$ by $A_Q\{2D\} =  \rho_0(b) v_2^2\{2D\}(b)$~\cite{davidhq}.

%The minimum-bias SS 2D jet peak in \pp collisions is strongly elongated in the {\em azimuth} direction, with approximate 2:1 aspect ratio~\cite{porter2,porter3}. The strong $\phi$ elongation in \pp collisions contrasts with strong {$\eta$} elongation in more-central \auau collisions, with 3:1 aspect ratio~\cite{axialci,daugherity}.

\subsection{Fluctuations from correlations}

The direct relation between fluctuations and two-particle angular correlations was established in Ref.~\cite{inverse} and implemented for $\langle p_t \rangle$ fluctuations in~\cite{ptscale,ptedep}. Given variance  $\sigma^2_{n} = \langle n^2 \rangle - \bar n^2$ the per-particle number fluctuation measure $\Delta \sigma^2_n(\delta x,\Delta x) \equiv (\sigma^2_n - \bar n) / \bar n$ is equivalent to the integral up to scale (bin size) $\delta x$ within some acceptance $\Delta x$ of per-particle number angular correlation measure $\Delta \rho / \sqrt{\rho_{ref}}$ defined on difference variable $x_\Delta$~\cite{inverse}.

The negative binomial distribution (NBD) is a two-parameter multiplicity distribution for some bin size $\delta x$ over some acceptance $\Delta x$ (which should be specified). NBD parameters are mean multiplicity $\bar n$ and parameter $k$ which can be interpreted as a number of independent particle sources. In the NBD context $\Delta \sigma^2_n(\delta x,\Delta x) = \bar n / k$ (correlated particles per independent source) may then represent the integral up to a specific bin size (scale) of  angular correlations arising from a superposition of several physical mechanisms within a specific acceptance. $\bar n/k$ typically increases monotonically with increasing bin size, angular acceptance and correlation amplitude.  Systematic details may reflect several underlying correlation mechanisms. Only differential correlation analysis can identify individual correlation sources (see Sec.~\ref{hadcor}).

%%%%%%%%%
 \section{Gluon Correlations from Glasma} \label{gluecorr}

Description of so-called ``bulk'' hadron production in terms of a CGC Glasma is an alternative to conventional parton distribution functions (PDFs) and longitudinal fragmentation of projectile nucleons. According to Ref.~\cite{lappi} bulk hadron production arises from low-$x$ gluons residing in the wavefunctions of projectile nucleons, and ``leading correlations are present already in the wave functions of the colliding objects.''

If a characteristic energy scale (saturation scale $Q_s$) is large enough $(Q_s \gg \Lambda_{QCD})$ the longitudinal system may be divided into radiating color charges at larger $x$ (sources) and a (nominally boost invariant) saturated classical field (Glasma) at smaller $x$. The transverse correlation length for the Glasma field is $1/Q_s$, interpreted as the transverse size of Glasma ``flux tubes.'' 

Some features of angular correlations in more-central \auau collisions, particularly $\eta$ elongation of the SS 2D peak, have been interpreted as a ``signal'' for Glasma flux tubes in the \aa initial state (IS)~\cite{lappi,gmm}. The present analysis is intended to test the conjecture that $\eta$ elongation of the  same-side 2D peak in more-central \auau collisions can be explained in terms of interaction of initial-state Glasma flux tubes with radial flow.

\subsection{Glasma gluon fluctuations and the NBD}

In the Glasma model $N_{FT} = Q_2^2 S_\perp  \gg 1$ is the number of flux tubes in an \aa system with transverse area $S_\perp$~\cite{lappi}, and $N_{FT} = O(1)\, N_{part}$~\cite{tomglasma}. 
Each flux tube emits into $N_c^2 - 1$ color states for a total of $(N_c^2 - 1) N_{FT} / 2\pi $ independent Bose-Einstein color radiators. Each radiator emits on average $\bar n_g = O(1) / \alpha_s(Q_s^2)$ gluons.
In Eq.~(3.2) of Ref.~\cite{lappi} the gluon total multiplicity is given by $\bar N_g = (f_N / \alpha_s)Q_s^2 S_\perp$.  

The Glasma model is thus a two-tiered statistical system. Fluctuations in the Glasma gluon multiplicity can be modeled by a negative binomial distribution with parameters $\bar N_g$ and $k$. 
According to the Glasma model in Ref.~\cite{lappi} multigluon correlations in the classical limit expressed in terms of the variance of gluon number $N_g$ 
%$N_g \sim 1/\alpha_s$ (number of gluons per quantum state $\sim$ flux tube) 
have  $\Delta \sigma^2_{N_g} \equiv (\sigma^2_{N_g} - \bar N_g) / \bar N_g = \bar N_g / k  = 0$ or no correlations. Averaging over \aa color source configurations induces significant correlations, with $\bar N_g / k \neq 0$.
%
%More generally, in Eq. 3-2 $m_q = \langle n_g^{(q)} \rangle - \bar n_g^q  = \bar n_g (q-1)! (\bar n_g / k)^{q-1}$ with $m_2/\bar n_g =  \bar n_g / k$ as above.
%\subsection{Glasma fluctuations}
Correlations in \pp collisions are expected to be small ($k \rightarrow \infty$). However, that trend is inconsistent with observations (Sec.~\ref{ppcorr}).
Glasma gluon fluctuations are further discussed in Sec.~\ref{fluctcorr}. 

%NBD parameter $k$ measuring non-Poisson gluon fluctuations is related to the model by parameter $\kappa_2 = N_{FT} / k$ with $\kappa_2 = O(1)$ defined in Eq.~(\ref{fluxtube}).  

\subsection{Glasma predicted two-gluon correlations}

Perturbative results from Ref.~\cite{lappi2} are summarized by
\bea \label{fluxtube}
C_2(\vec p_1,\vec p_2) \rightarrow \Delta \rho(\vec p_1,\vec p_2) &=& \rho(\vec p_1,\vec p_2)- \rho_{ref}(\vec p_1,\vec p_2) \nonumber \\\nonumber
\frac{\Delta \rho(\vec p_1,\vec p_2)}{\rho_{ref}(\vec p_1,\vec p_2)}&=& \frac{1}{k} =  \frac{\kappa_2}{Q_s^2 S_\perp}  \\ \nonumber
\rho_{ref}(\vec p_1,\vec p_2) &=& \rho_0(\vec p_1)\rho_0(\vec p_2) \nonumber \\
\rho_0(\vec p) &\propto& \ln(p_t / Q_s)  (Q_s / p_t)^4,
\eea
with NBD parameter $k \approx (N_c^2 - 1)N_{FT}/2\pi$~\cite{lappi} and $\kappa_2 = N_{FT} / k  \approx  2\pi / (N_c^2 - 1) = O(1)$ approximately constant for $p_t / Q_s \gg 1$~\cite{lappi2}. Since $k \approx N_{FT}$ we have $N_g \approx n_g N_{FT}$. The Glasma predicted energy dependence is
$k \sim Q_s^2 \sim \sqrt{s}^\lambda$~\cite{lappi}.
As defined, $\kappa_2$ has the per-particle~\cite{axialci,daugherity,inverse} structure $\bar n_{ch} (\hat r-1) \rightarrow N_{FT} (\hat r-1)$. 
For the perturbative case the only correlation source is production hierarchy: multiple gluons associated with each independent flux-tube radiator. Perturbative two-gluon correlations are then independent of angle differences and factorizable on $(p_{t1},p_{t2})$. A non-perturbative calculation is required to determine correlations down to small $p_t$~\cite{lappi2}. $\kappa_2$ may then depend on $p_t$ and relative angle $\phi_\Delta$, but not on $\eta_\Delta$ if flux tubes are boost invariant. 

%\bea \label{fluxtube}
%\kappa_2(\vec p_{t1},\vec p_{t2}) = Q_2^2 S_\perp \Delta \rho(\vec p_{t1},\vec p_{t2}) / \rho_{ref}(\vec p_{t1},\vec p_{t2}),
%\eea

%Gavin adjusted $\kappa_2$ to fit data. Here is attempted an absolute prediction. 

%The Glasma gluon spectrum peaks well below 1 GeV/c.

% $k \approx (N_c^2 - 1) Q_s^2 S_\perp / 2\pi$ NBD constant

%%%%%%%%%%%%%%
\section{Measured hadron correlations} \label{hadcor}

There is now an extensive phenomenology of hadron correlations from which the underlying parton dynamics may be inferred. The phenomenology described here exhausts all differential correlation structure in \pp and \auau collisions, and therefore all fluctuation phenomenology. Fluctuation measures, as running integrals of differential correlations, are consistent with correlation measurements but retain less information.

%Similarly,  correlations plus their integrals as measured by fluctuations, spectra and yields.

\subsection{Correlations from \pp collisions} \label{ppcorr}

%The PYTHIA description of \pp collisions is essentially a two-component model.

\begin{figure}[h]
   \includegraphics[width=1.65in,height=1.5in]{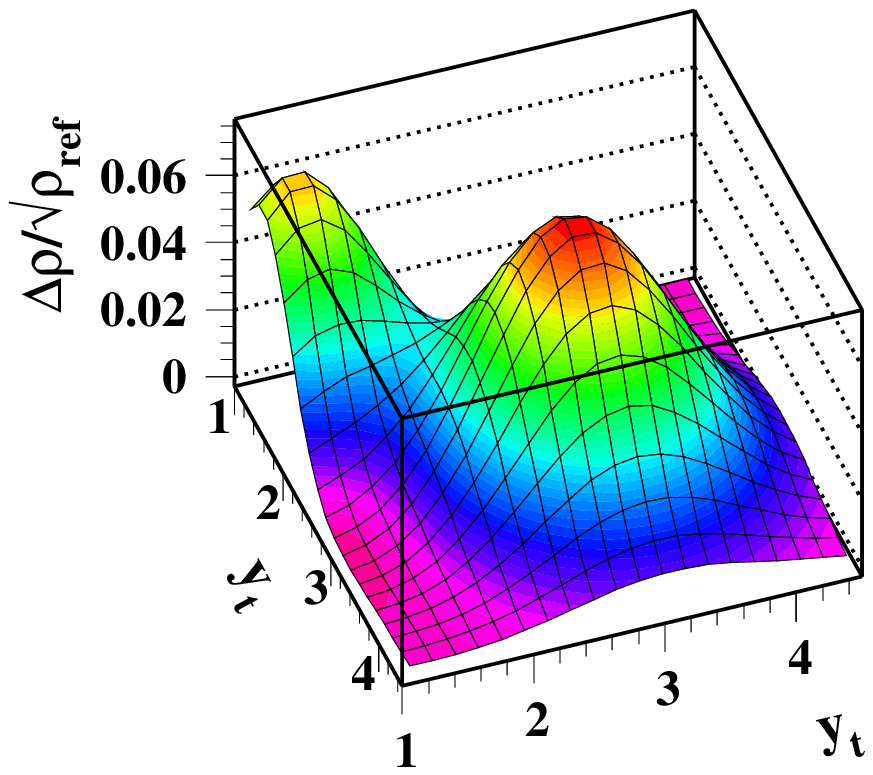}
 \includegraphics[width=1.65in,height=1.5in]{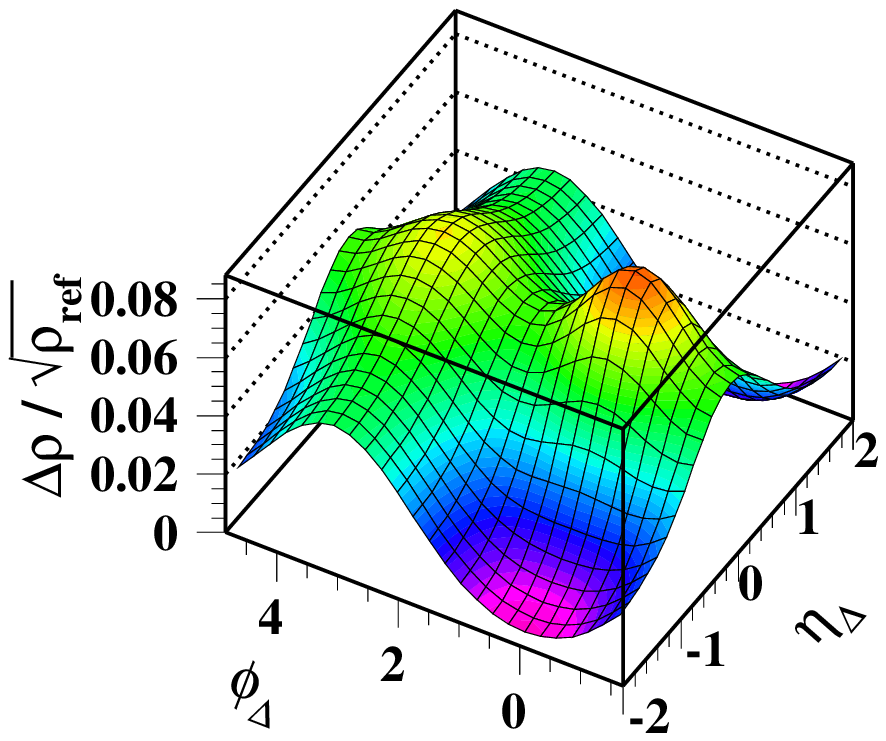}
 \caption{\label{ppcorr1} (Color online)
Two-particle hadron correlation histograms from 200 GeV \pp collisions based on data parametrizations from Refs.~\cite{porter1,porter2}. Left: $y_t \times y_t$ correlations within the angular acceptance showing soft and hard components, Right: $p_t$-integral angular correlations on $(\eta_\Delta,\phi_\Delta)$ showing a superposition of soft and hard components.
 } % ppcmsx1a, 1b
 \end{figure}

Figure~\ref{ppcorr1} shows parametrizations of \ytyt and 2D angular correlations from \pp collisions at 200 GeV~\cite{porter2,porter3}. Charged-hadron correlations on \ytyt (left panel) with $y_t \in [1,4.5]$ or $p_t \in [0.15,6]$ GeV/c integrated over the STAR TPC angular acceptance $|\eta|<1$ and $2\pi$ azimuth include two well-separated peaked structures described as the soft component and the hard component~\cite{ppprd,porter2}. 
The soft component is a 2D peak localized below 0.5 GeV/c (e.g.\ within $y_{t1} + y_{t2} < 2$). The hard component is a second distinct 2D peak which dominates the complementary \ytyt space. The peak is centered near $y_t \approx 2.7$ ($p_t \approx 1$ GeV/c) and does not extend below 0.35 GeV/c in \pp collisions. Corresponding structures are superposed in minimum-bias angular correlations (right panel).

%%%%%%%%%%
%%%%%%%%%%%%

\begin{figure}[h]
  \includegraphics[width=1.65in,height=1.5in]{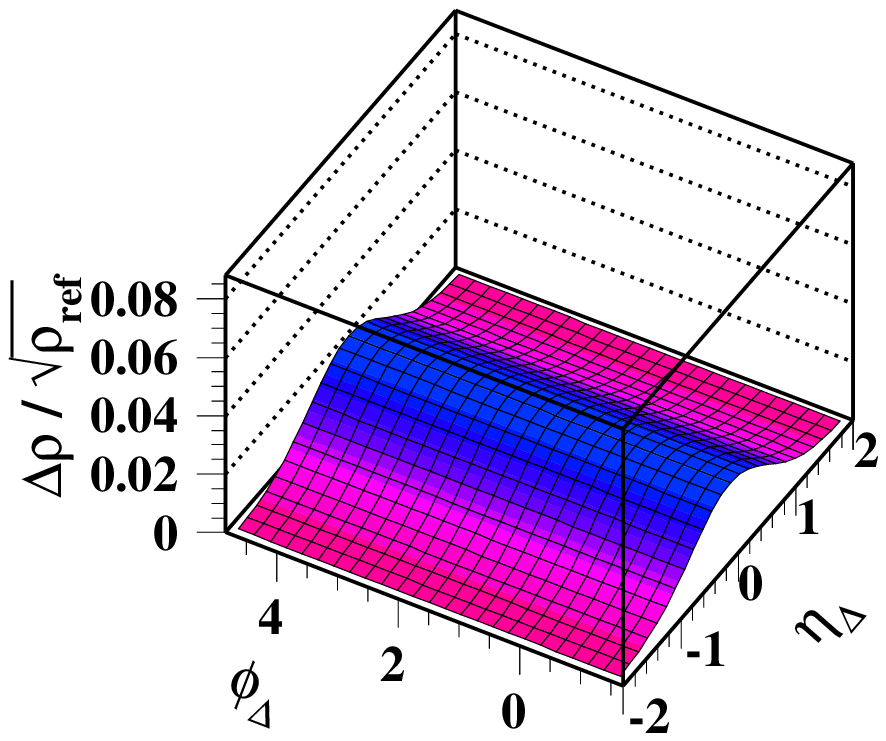}
  \includegraphics[width=1.65in,height=1.5in]{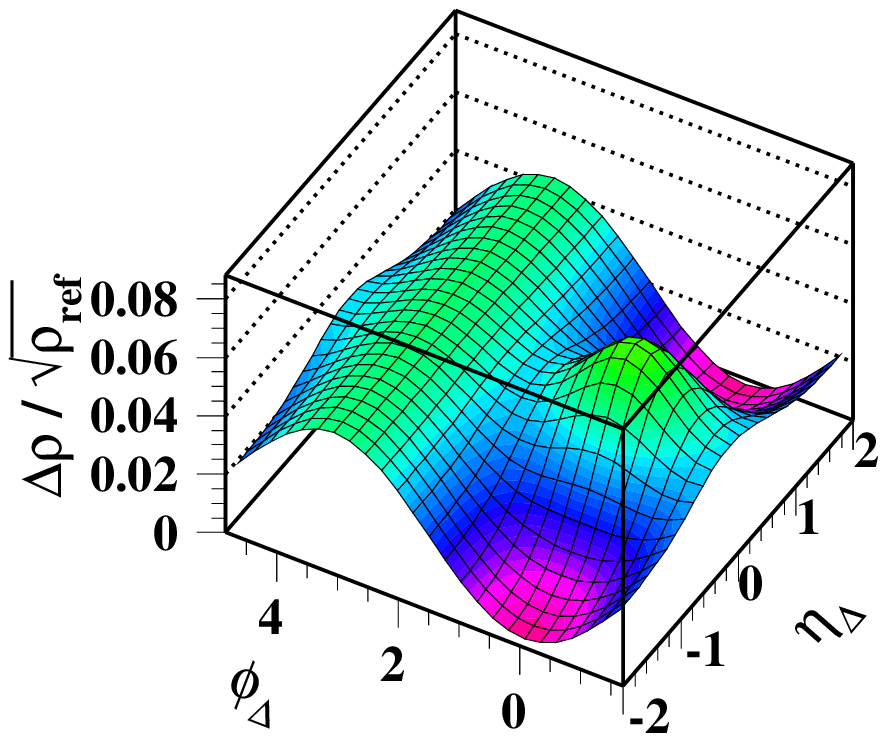}
\caption{\label{ppcorr2} (Color online)
Two-particle hadron correlation histograms from 200 GeV \pp collisions based on data parametrizations from Refs.~\cite{porter1,porter2}. Left: Soft component (longitudinal projectile nucleon fragmentation to US pairs), Right: Hard component (large-angle-scattered parton fragmentation to US SS pairs and CI=LS+US AS pairs).
 } % ppcmsx1a, 1b
 \end{figure}

Figure~\ref{ppcorr2} shows distinct structures in angular correlations on $(\eta_\Delta,\phi_\Delta)$ corresponding to the soft and hard components separated by cuts on \ptpt or $y_t\times y_t$. The soft component of angular correlations (left panel) is a 1D Gaussian on $\eta_\Delta$ with approximately unit r.m.s. width characteristic of ``short-range'' correlations. The soft component is almost entirely US pairs. Such structure is consistent with diffractive scattering and longitudinal fragmentation of projectile nucleons to hadrons (near mid-rapidity) in charge-neutral (US) combinations.

%%%%%%%%%%
%%%%%%%%%%%%

The hard-component peak on $(y_t,y_t)$ corresponds to two structures on $(\eta_\Delta,\phi_\Delta)$ (right panel), a SS 2D peak centered at the angular difference origin and an AS ridge centered at $\pi$ on $\phi_\Delta$ and uniform on $\eta_\Delta$. The SS 2D peak is almost entirely US pairs (reflecting local charge conservation during fragmentation), whereas the AS 1D peak on azimuth (jet-jet ridge) is composed of equal numbers of LS and US pairs (no interjet charge correlation). 

The structure on $(y_t,y_t)$ corresponding to the SS 2D peak lies close to the main diagonal ($y_{t1}\approx y_{t2}$), whereas that corresponding to the AS ridge is substantially broadened relative to the main diagonal~\cite{porter2,porter3}.  Those systematics are consistent with large-angle parton (gluon) scattering and fragmentation to back-to-back jet pairs, with hadron local charge and momentum conservation.

\subsection{Correlation evolution with \aa centrality}

If \aa collisions were simply linear superpositions of \pp (or N-N) collisions (eikonal model) we could extrapolate \pp correlation phenomenology according to the Glauber model to describe \aa data. The eikonal model provides a {reference system}  denoted by {\em Glauber linear superposition} (GLS), including participant scaling of the soft component and binary-collision scaling of the hard component~\cite{kn,hardspec}. Correlation data from \auau collisions at 62 and 200 GeV follow the GLS reference from peripheral collisions to an intermediate centrality and then transition to substantially different behavior~\cite{daugherity}.

%%%%%%%%%%
 \begin{figure}[h]
  \includegraphics[width=1.65in,height=1.5in]{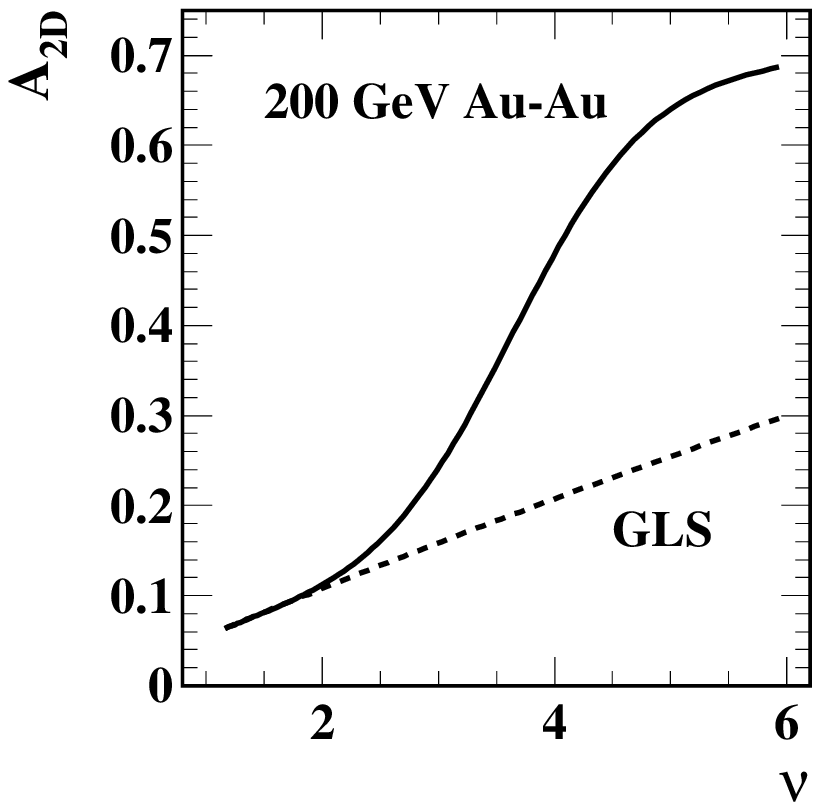}
  \includegraphics[width=1.65in,height=1.53in]{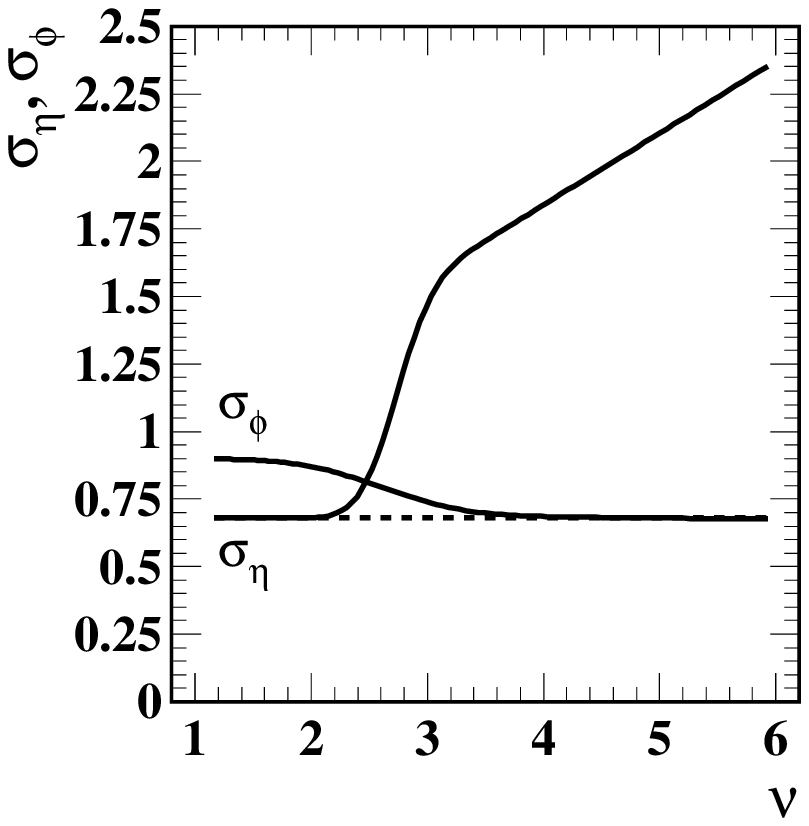}
\caption{\label{fitparams}
Left: Amplitude of the same-side 2D Gaussian fitted to minimum-bias 2D angular correlation data from 200 GeV \auau collisions~\cite{daugherity}.
Right: Fitted peak widths for the same-side 2D Gaussian. GLS indicates a Glauber linear superposition reference extrapolated from measured \pp collisions~\cite{ppprd}.
 } 
 \end{figure}
%%%%%%%%%%%%

Figure~\ref{fitparams} summarizes centrality variation of SS 2D peak parameters. Deviations from the GLS reference extrapolation above the transition point include a rapid change (within one 10\% centrality bin) in the rate of increase (slope on centrality measure $\nu$ defined in Sec.~\ref{ppcent}) of SS and AS jet-related peak amplitudes, a rapid increase in the SS 2D peak $\eta_\Delta$ width and a significant {\em decrease} in the $\phi_\Delta$ width.  The SS 2D peak aspect ratio transitions from nearly 2:1 elongation on azimuth to 3:1 elongation on pseudorapidity~\cite{porter3}. Jets in \pp collisions are nearly as anomalous as those in central \auau collisions. Those trends have been interpreted quantitatively in terms of modified parton fragmentation in more-central \auau collisions~\cite{fragevo}.

%%%%%%%%%%
 \begin{figure}[h]
  \includegraphics[width=1.65in,height=1.6in]{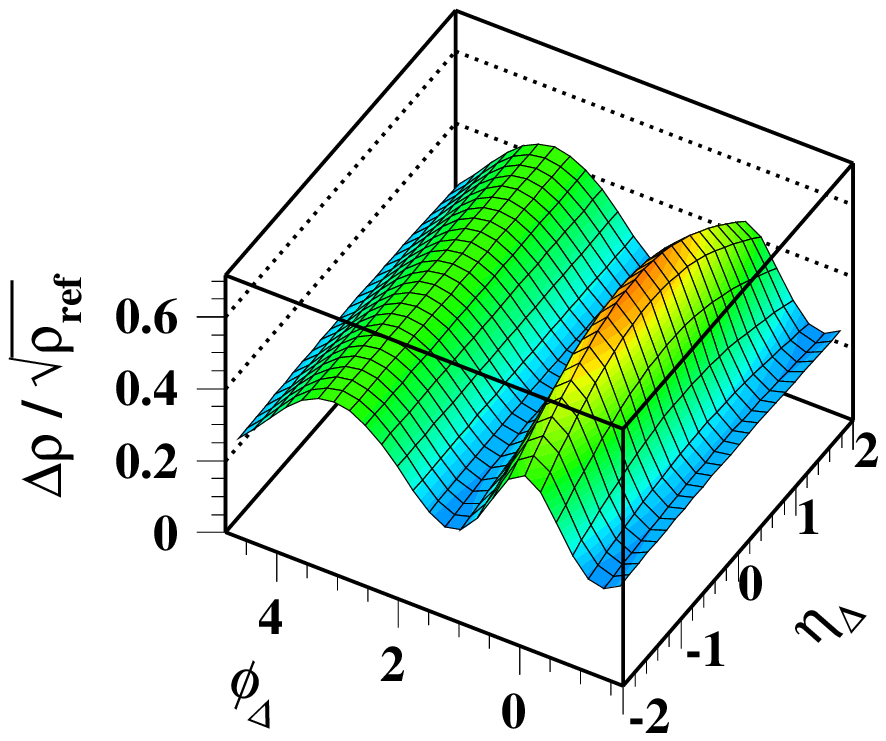}
  \includegraphics[width=1.65in,height=1.6in]{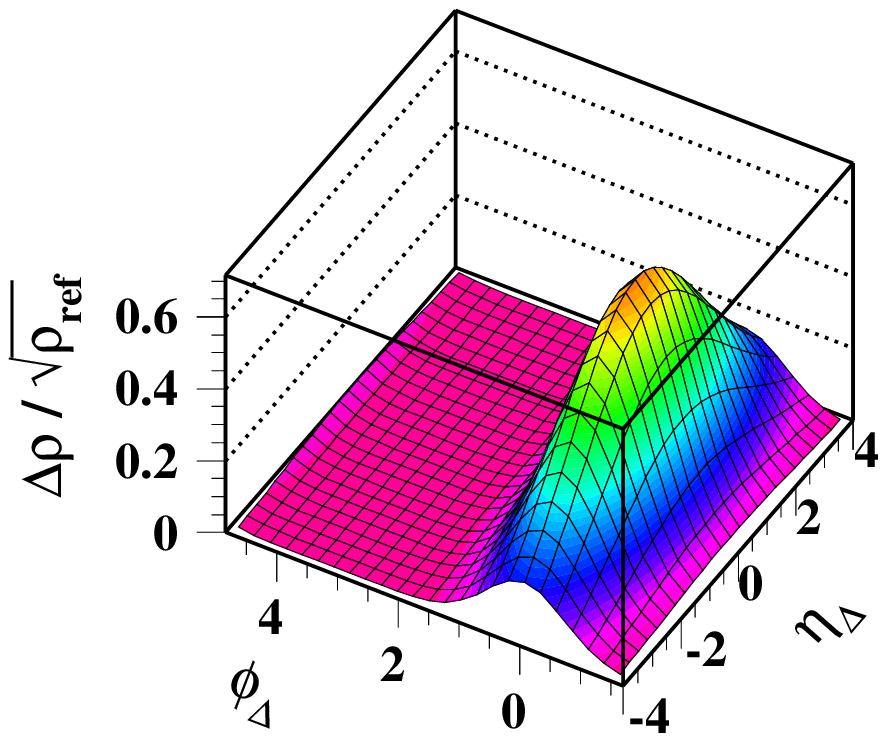}
\caption{\label{jetpeaks} (Color online)
Left: Angular correlations from 200 GeV \auau collisions with centrality $\nu = 6 ~(b=0)$ with nonjet fit components (quadrupole, 1D Gaussian on $\eta_\Delta$) subtracted to reveal nominal jet correlations. The vertical zero is the estimated true zero offset for these overlapping jet correlations.
Right: The previous histogram with the away-side dipole term subtracted to isolate the same-side 2D jet peak extrapolated to $\eta \in [-2,2]$.
 } % ppcmsx1a, 1b
 \end{figure}
%%%%%%%%%%%%

Figure~\ref{jetpeaks} shows jet structure in 200 GeV central ($b = 0$) \auau collisions extrapolated from measured centrality trends. The nonjet quadrupole measured by $v_2\{2D\}$ is zero with small upper limit in that case~\cite{davidhq}. The AS 1D peak on azimuth is uniform on $\eta_\Delta$ within the STAR TPC acceptance, as is the nonjet quadrupole. In Fig.~\ref{jetpeaks} (right panel) the $\eta_\Delta$-independent components have been subtracted, leaving the SS 2D peak (extrapolated to $|\eta_\Delta| = 4$) as the remaining correlation component.
The SS 2D peak for {\em minimum-bias} ($p_t$-integral) data is always consistent with a 2D Gaussian (no additional structure). Although ``trigger-associated'' dihadron correlations do reveal non-Gaussian (on $\eta_\Delta$) SS features, those data are subsets of the minimum-bias ensemble and should reflect the same basic fragmentation  process.

\subsection{Relation to hadron spectra and yields}

Soft and hard components of two-particle correlations have counterparts in single-particle hadron $p_t$ spectra. The \pp $p_t$ spectrum hard component~\cite{ppprd} corresponds quantitatively to the hard component in Fig.~\ref{ppcorr1} (left panel). \pp spectra and correlations form a simple system quantitatively consistent with all aspects of pQCD down to zero hadron momentum and 3 GeV parton energy~\cite{fragevo}.

In a minijet context the $p_t$ spectrum hard component (single-particle fragment distribution) is the marginal projection of the  correlation hard component on \ptpt or \ytyt (fragment pair distribution~\cite{porter2}). For all \auau centralities the measured correlation hard component on \ptpt\cite{lanny} is quantitatively consistent with the hard component inferred from measured $p_t$ or $y_t$ spectra~\cite{hardspec}. The hard-component peak on \ptpt or \ytyt persists as a {\em distinct structure} with mode near $p_t = 1$ GeV/c even in central \auau collisions. For both spectra and correlations, suppression at larger $p_t \sim 10$ GeV/c is accompanied by much larger enhancement at smaller $p_t \sim 0.5$ GeV/c~\cite{hardspec}. Suppression and enhancement trends on \auau centrality are closely (anti)correlated~\cite{fragevo}.

The integral of the SS 2D peak in angular correlations (all jet-related hadron pairs) combined with a pQCD dijet cross section can be converted to a hard-component hadron yield. Variation of the calculated hard-component yield with centrality explains evolution of the \auau total hadron yield with centrality, revealing that about one third of the total yield in central \auau collisions is included in resolved minijets~\cite{jetspec}.
Analysis of the $p_t$ or $y_t$ dependence of the SS 2D $\eta$-elongated peak in angular correlations~\cite{davidhq2,davidaustin,lanny} shows that this structure corresponds to the hard-component peak on \ptpt or \ytyt near $p_t = 1$ GeV/c for all \auau centralities.

%With increasing \auau centrality the $p_t$ spectrum and \ptpt correlation hard components do exhibit substantial evolution which remains however compatible with a pQCD description. 

%That system of inferred gluon correlations can be compared with predictions from Glasma flux-tube arguments.

%On ytxyt AS we have a two-gluon distribution with kt broadening of two kinds again folded with two-particle AS FF (interjet, think about that), but this time no CD component.

%On ytxyt SS we might have the self-pair parton spectrum folded with two-particle SS FF (intrajet) to give SS ytxyt hard component. Then soft component with US pairs reveals a very soft parent distribution, in terms of $p_t$.

%%%%%%%%%
\section{Glasma flux tubes ${\bf vs}$ hadron data} \label{glasmavs}

% QUESTIONING RADIAL FLOW AND NONHYDRO

The Glasma model is a one-component (soft) model of gluon production near mid-rapidity emphasizing more-central \aa collisions. There is no (semi)hard parton (gluon) scattering to mid-rapidity. The Glasma model competes with the two-component (soft plus hard) model of hadron production inferred from yields, spectra and two-particle correlations derived from \pp and \aa collisions~\cite{ppprd,hardspec,porter2,porter3,axialci,daugherity}. We provide direct comparisons between Glasma predictions and measured hadron spectrum and correlation data as a test of Glasma relevance to the hadronic final state in nuclear collisions. We consider spectrum and correlation trends on $(\eta,\phi)$, $p_t$ or $y_t$, \aa centrality and collision energy.

%We want to know what is the Glasma prediction for momentum, b and root-s dependence of $\kappa_2$ as the correlation measure of Glasma.

\subsection{Glasma single-gluon spectrum}

The number of Glasma flux tubes is expected to scale with \aa centrality as $Q_s^2 S_\perp \approx N_{part}$~\cite{tomglasma}. The corresponding gluon production should then be compared with the soft component of hadron production which does follow $N_{part}$ scaling~\cite{ppprd,hardspec}.
The inclusive single-gluon spectrum from Ref.~\cite{lappi2}, Fig.~3 (left panel) is well represented by 
\bea \label{glaseq}
\frac{dN_g}{dp_t^2} \hspace{-.0in} &=& \hspace{-.0in} \text{erf}(p_t / \text{0.3 GeV/c}) \frac{1/p_t^2}{\{(1/5)^2 + (p_t^2 / 33)^2\}^{\frac{1}{2}}},
\eea
whereas the soft-component charged-hadron spectrum for \pp and all centralities of \aa is the L\'evy distribution
\bea \label{hadreq}
\frac{2}{n_{part}}\frac{d^2n_{ch}}{d\eta\, m_t dm_t} &=&  \frac{45}{\{1 + (m_t - m_\pi) / n\, T\}^n},
\eea
with $T = 0.145$ GeV and $n = 12.8$~\cite{ppprd,hardspec}.

%%%%%%%%%%
 \begin{figure}[h]
  \includegraphics[width=1.65in,height=1.5in]{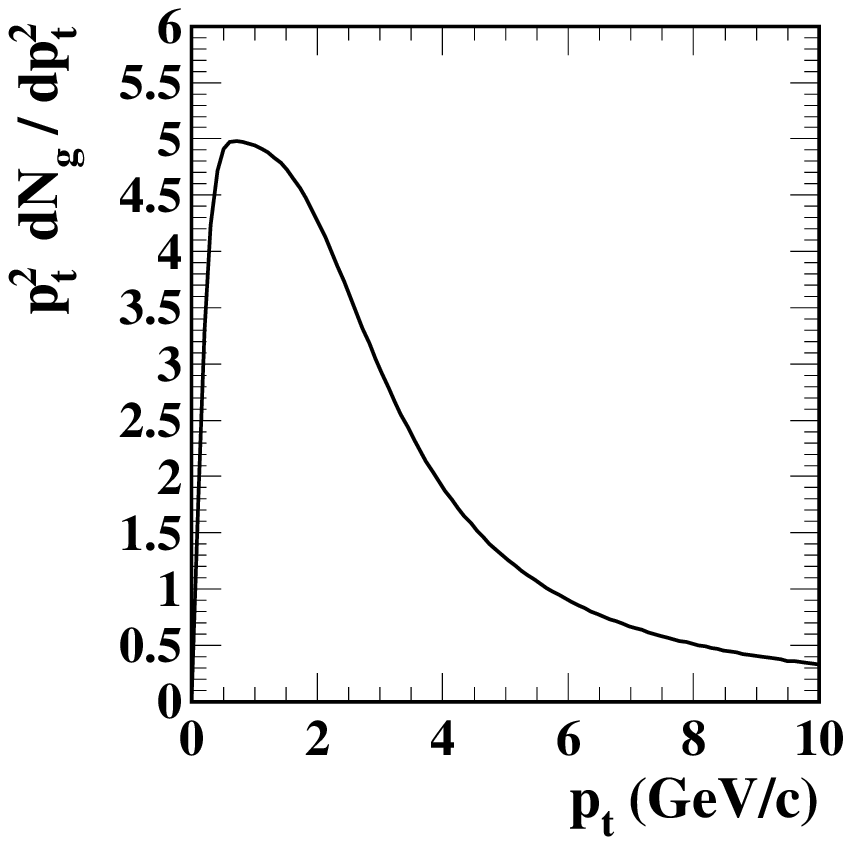}
   \includegraphics[width=1.65in,height=1.5in]{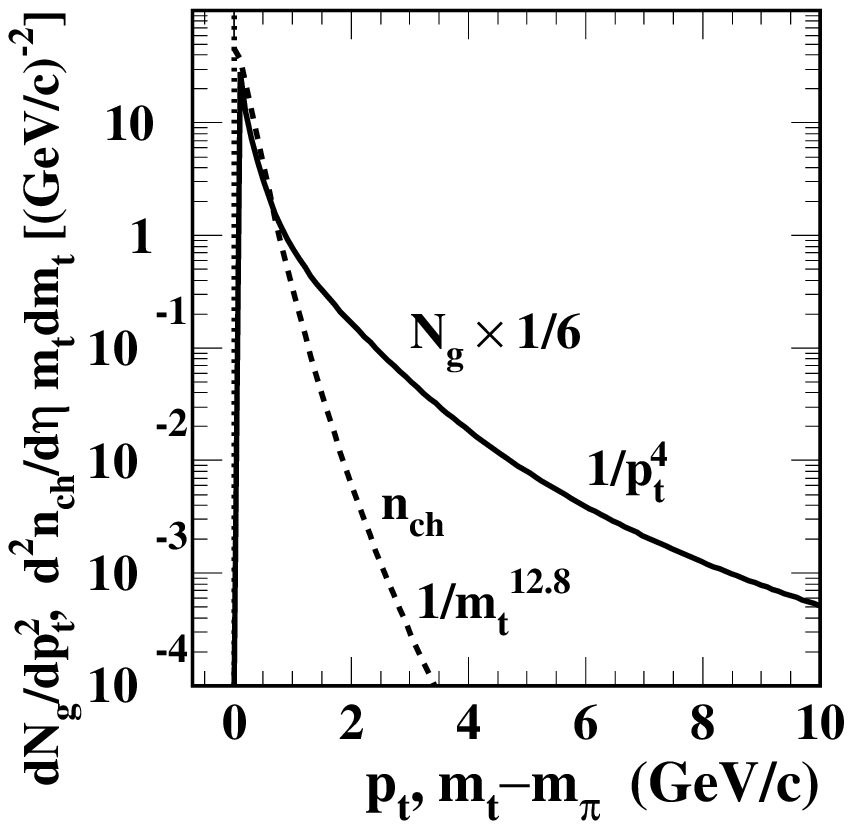}
\caption{\label{glasspec} 
Left: Gluon $p_t$ spectrum from Glasma flux-tube model (with added factor $p_t^2$).
Right: Comparison of hadron $m_t$ spectrum soft component from Ref.~\cite{ppprd} (dashed curve) and gluon $p_t$ spectrum from Glasma flux-tube model (solid curve).
 } 
 \end{figure}
%%%%%%%%%%%%

Figure~\ref{glasspec} (left panel) shows Eq.~(\ref{glaseq}) which compares well with Fig.~3 (left panel) from Ref.~\cite{lappi2} where the spectrum has been scaled by factor $p_t^2$ to show details at smaller $p_t$. The Glasma expectation for the gluon spectrum trend is $\propto \ln(p_t  / Q_s)(Q_s / p_t)^4$ at larger $p_t / Q_s$ with $Q_s \approx 1$ GeV~\cite{lappi2}.

Figure~\ref{glasspec} (right panel) shows a comparison between the Glasma gluon spectrum $\propto 1/p_t^4$ and the per-participant-pair hadron spectrum soft component $\propto 1/m_t^{12.8}$ at larger $p_t$ and $m_t$ respectively.  The $m_t$ spectrum integrates to $dn_{ch}/d\eta = 2.5$ (NSD \pp collisions). The gluon spectrum in Eq.~(\ref{glaseq}) has been divided by 6 to approximate the same $\eta$ density for the shape comparison. The hard gluon spectrum has no observed $N_{part}$-scaling counterpart in the hadronic final state. There is no correspondence in the gluon spectrum to the measured hard component scaling as $N_{bin}$ in hadron spectra~\cite{ppprd,hardspec} or to the large-angle-scattered parton spectrum varying as $1/p_t^{6.5}$~\cite{fragevo}.

\subsection{Glasma and NBD $\bf k$-parameter systematics} \label{gluespec}

%As an integral correlation measure the NBD $k$ parameter depends on details of collision system size, detector acceptance and analysis bin size. 

\begin{figure}[h]
  \includegraphics[width=1.65in,height=1.5in]{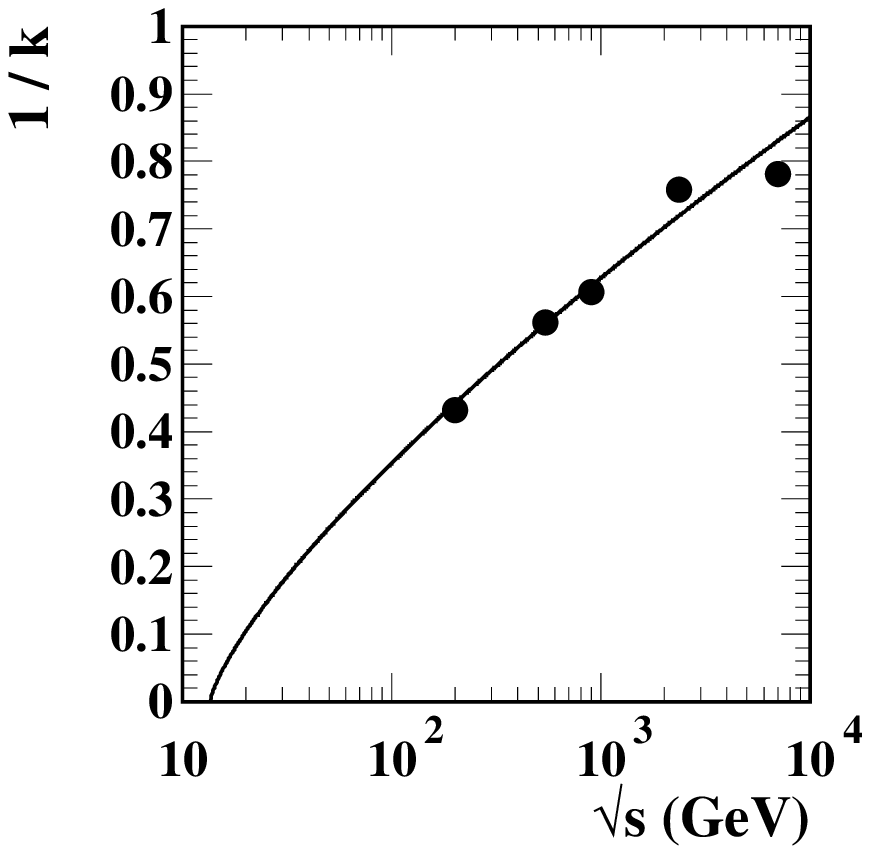}
   \includegraphics[width=1.65in,height=1.5in]{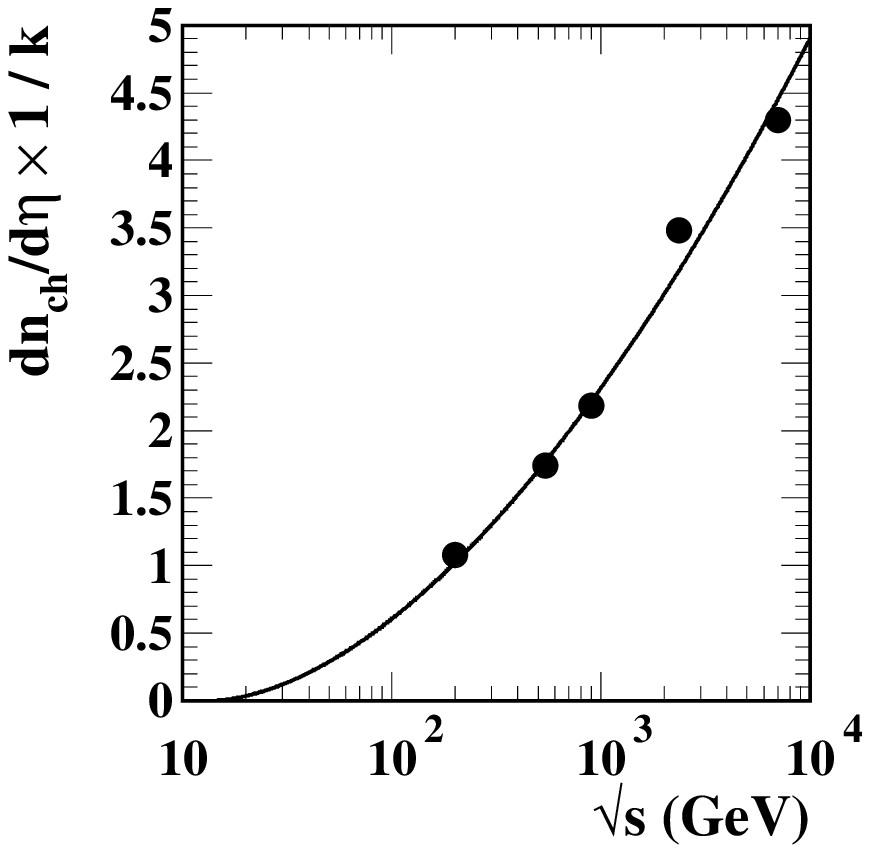}
\caption{\label{fig1} Left: Energy dependence of negative binomial distribution (NBD) parameter $k$ from Ref.~\cite{lappi} plotted as $1/k$ for \pp collisions and angular acceptance $|\eta| < 1$ or $\Delta \eta = 2$, Right: The same data plotted as $dn_{ch}/d\eta \times 1/k$ measuring the integral of angular correlations within the anglar acceptance. The curves are based on $\ln(\sqrt{s} / \text{13.5 GeV})$ (see text).
 }  %left panel is $\propto \kappa_2$
 \end{figure}

In Fig.~\ref{fig1} (left panel) we show $k$ data from Fig. 3 of Ref.~\cite{lappi} plotted
 as $1/k$, which then increase by factor 2 from RHIC to LHC energies. For ``short-range'' (localized on $\eta$) angular correlations we do expect $1/k$ (representing {\em per-pair} angular correlations) to decrease with increasing {\em system size} (e.g.\ \aa centrality). In the CGC model the $k \sim Q_s^2$ parameter is expected to increase monotonically with collision energy, or equivalently $1/k$ should decrease with collision energy. 
The observed strong $1/k$ increase with energy contradicts the Glasma expectation.

%%%%%%%%%%
%%%%%%%%%%%%

In Fig.~\ref{fig1} (right panel) corresponding product $dn_{ch} /d\eta \times 1/ k$ measures {\em per-particle} angular correlations integrated on scale up to the angular acceptance (in this case $\Delta \eta = 2$ and $\Delta \phi = 2\pi$). We conclude that integrated angular correlations in \pp collisions from whatever mechanism increase with collision energy faster than $\log(\sqrt{s})$.

Given the measured energy dependence of $p_t$ angular correlations~\cite{ptedep}, nonjet azimuth quadrupole correlations~\cite{davidhq} and minijet angular correlations~\cite{daugherity} in \pp and \auau collisions at and below 200 GeV we characterize the $k$ data based on the energy trend $\ln(\sqrt{s} / \text{13.5 GeV})$. The solid curve in the right panel is $dn_{ch} / d\eta \times 1/k = 0.18\{\ln(\sqrt{s} / \text{13.5 GeV})\}^{1.75}$. The hadron yield increase with energy is well described above 200 GeV by $dn_{ch} / d\eta \approx 0.88\{\ln(\sqrt{s} / \text{13.5 GeV})\}$. The solid curve in the left panel is therefore $1/k = 0.21\{\ln(\sqrt{s} / \text{13.5 GeV})\}^{0.75}$, generally consistent with QCD processes (e.g.\ minijet production) but inconsistent with the flux-tube expectation $k \propto \sqrt{s}^\lambda$ with $\lambda > 0$~\cite{lappi}. 

We can express the energy and centrality dependence of Glasma correlation parameter $\kappa_2$ in terms of measured quantities. Since $\kappa_2 = N_{FT} / k \approx N_{part} / k$ we have $\kappa_2 \approx (N_{part} /2 n_{ch}) \times (2n_{ch} / k$). Combining the two-component expression for the first factor~\cite{kn} with the trend inferred from Fig.~\ref{fig1} (right panel) we obtain $\kappa_2 \approx 0.36\{\ln(\sqrt{s} / \text{13.5 GeV})\}^{1.75} / \{1 + 0.1(\nu - 1)\}$: substantial increase with energy and decrease with centrality.

%Note the Froissart bound here. 

%What angular acceptance is assumed for the flux tube calculations? It seems to be $|\eta| < 1$. CHECK.

%This may include geometry/centrality fluctuations in p-p collisions, as well as the angular correlation contribution.

 \subsection{Glasma gluon  {$\bf p_t \times p_t$} correlations}

\begin{figure}[h]
  \includegraphics[width=1.65in,height=1.5in]{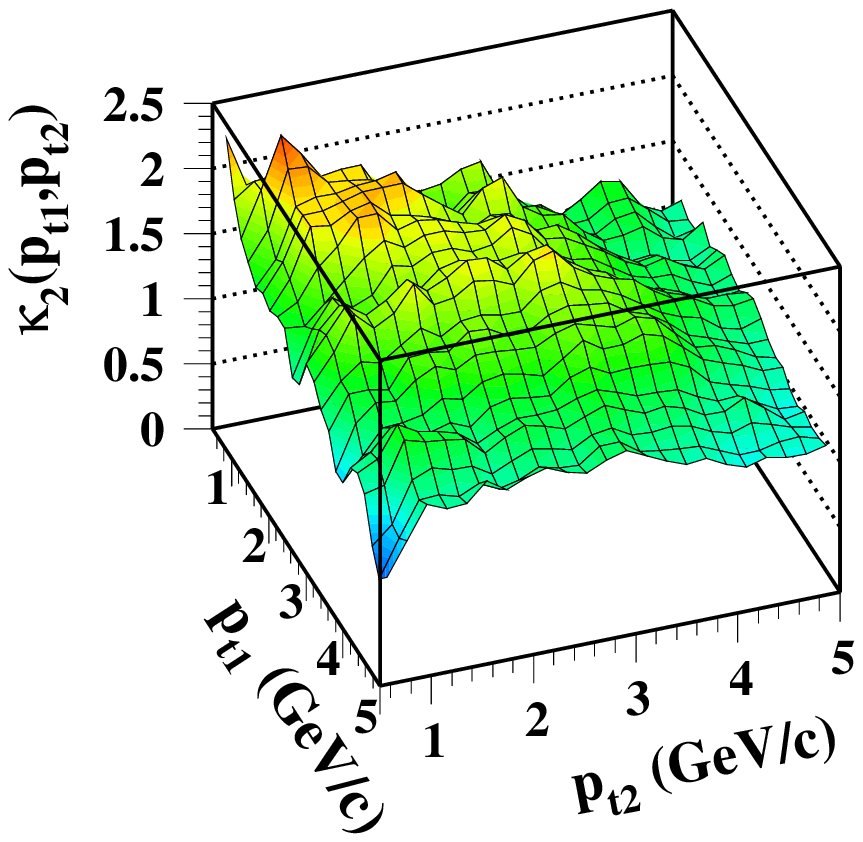}
  \includegraphics[width=1.65in,height=1.5in]{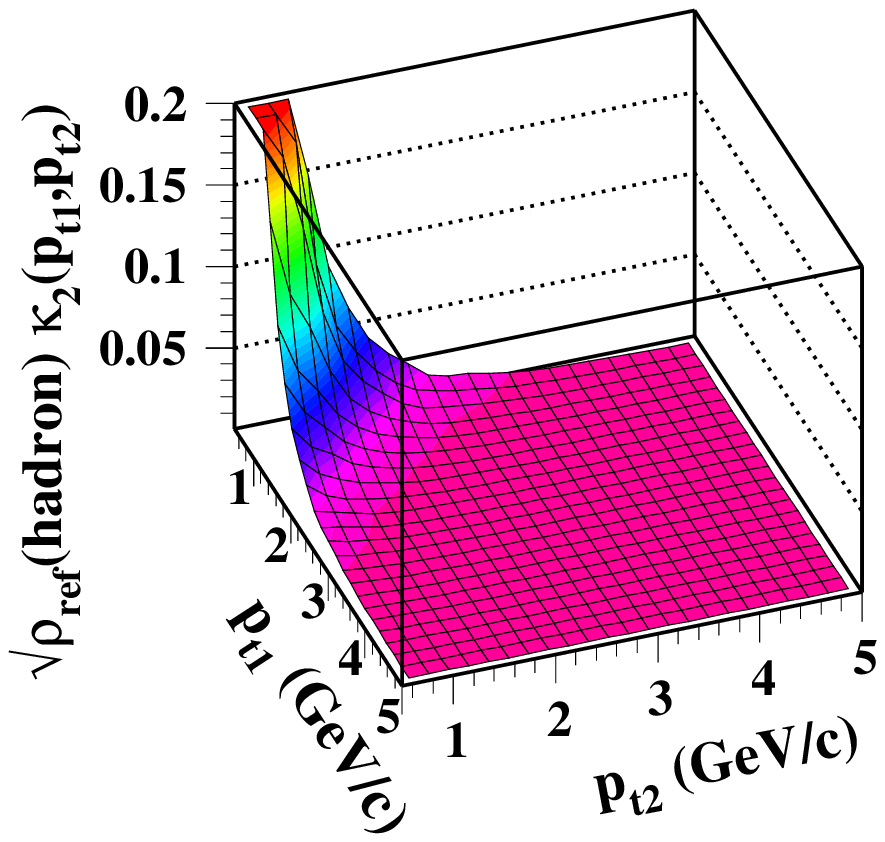}
\caption{\label{ptpt1} (Color online)
Left: Data from Fig.\ 2 (right panel) of Ref.~\cite{lappi} plotted in isometric view showing predicted Glasma two-gluon correlations.
Right: The same data with prefactor $\sqrt{\rho_{ref}(p_{t1},p_{t2})}$ based on a hadron spectrum soft component. 
}
 \end{figure}

Figure~\ref{ptpt1} (left panel) shows Fig.\ 2 (right panel, AS pairs) of Ref.~\cite{lappi}. Fig.\ 2 (left panel, SS pairs) of Ref.~\cite{lappi} includes self pairs along the diagonal but is otherwise statistically equivalent to the right panel. 
The cited figure shows a plan view of $\kappa_2$ on $(p_{t1},p_{t2})$ relative to saturation scale $Q_s \approx 1$ GeV/c. The histogram is replotted here in isometric view on $p_t$ in units GeV/c. 

Fig.~\ref{ptpt1} (left panel) is analogous to correlations plotted on $y_t \times y_t$ (Fig.~\ref{ppcorr1}, left panel). $\kappa_2$ prefactor $Q_s^2 S_\perp$ in Eq.\ (5.1) of Ref.~\cite{lappi} is equivalent to $\bar n_{ch}$ in $\bar n_{ch} \{\hat r(p_{t1},p_{t2}) - 1\}$ in the notation of Sec.~\ref{corrmeas}.  Thus, $\kappa_2$ should be compared with per-pair measure $\Delta \rho/\rho_{ref}(y_t,y_t)$~\cite{porter0} which increases dramatically at larger $y_t$, strongly contradicting the Glasma prediction plotted in Fig.~\ref{ptpt1} (left panel).  

%The per-pair measure is missing an essential prefactor leading to the large excursions at large $p_t$. 

%%%%%%%%%%
%%%%%%%%%%%%

Figure~\ref{ptpt1} (right panel) shows the $\kappa_2$ histogram in the left panel multiplied by prefactor  $ \sqrt{\rho_{ref}(p_{t1},p_{t2})} = \sqrt { \rho_0(p_{t1}) \rho_0(p_{t2})}$, where $\rho_0(p_{t})$ is the hadron soft component in Eq.~(\ref{hadreq}) divided by $2\pi$ to form a 3D density. 
Prefactor $\rho_0(b) = \bar n_{ch} / \Delta  \eta \Delta \phi$ appropriate for angular correlations is replaced by the geometric mean of  single-particle $p_t$ or $y_t$ spectra 
%$\sqrt { \rho_0(p_{t1},b) \rho_0(p_{t2},b)}$ 
appropriate for \ptpt correlations.  
The plotted histogram is proportional to per-particle measure $\Delta \rho/\sqrt{\rho_{ref}}(p_{t1},p_{t2})$ and thus directly comparable with per-particle correlation data. The result is consistent with measured soft-component \ptpt correlations.

%boosted by the maximum radial speed $\beta_t = 0.6$ inferred from central \auau collisions~\cite{radflow}. 

\begin{figure}[h]
  \includegraphics[width=1.65in,height=1.5in]{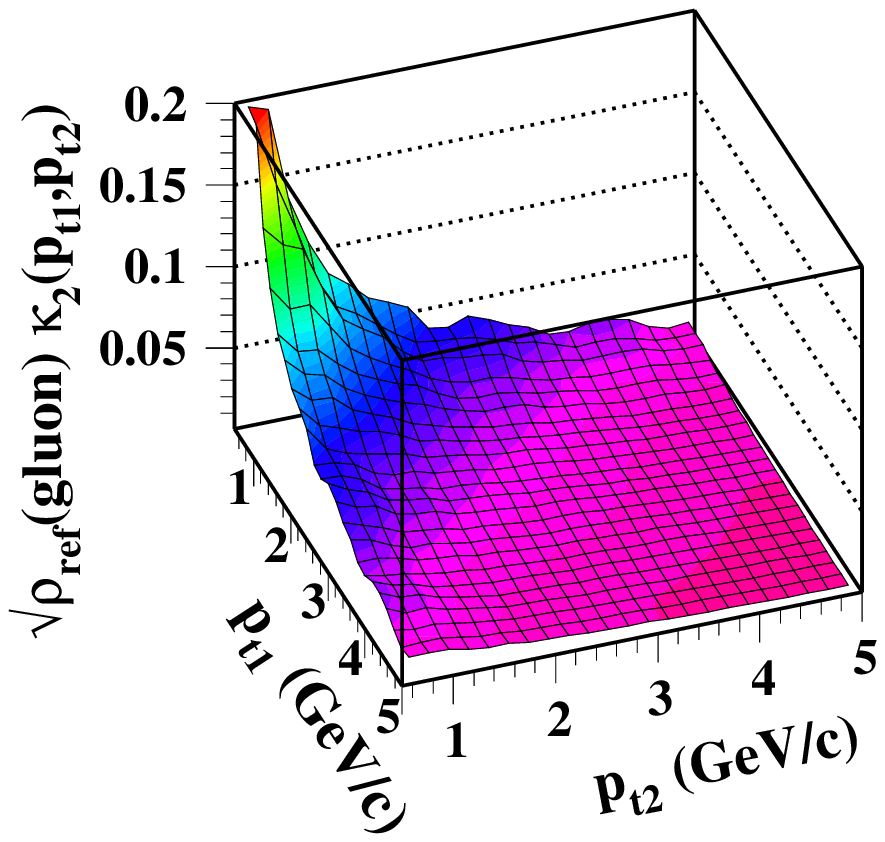}
  \includegraphics[width=1.65in,height=1.5in]{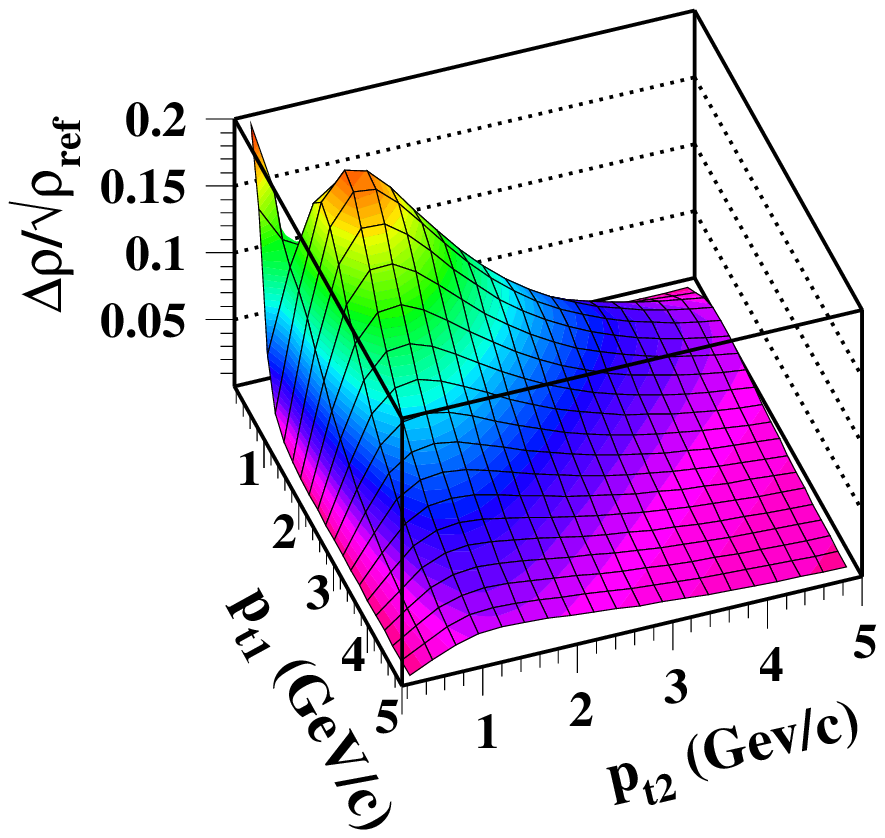}
\caption{\label{ptpt2} (Color online)
Left: Glasma predictions in the form $\Delta \rho / \rho_{ref}$ from Fig.~\ref{ptpt1} (left panel) multiplied by prefactor $\sqrt{\rho_{ref}} = \sqrt{\rho_0(p_{t1},b) \rho_0(p_{t2},b)}$ (gluon spectra) to obtain the per-particle form  $\Delta \rho / \sqrt{\rho_{ref}}$.
Right: The $y_t \times y_t$ histogram in Fig.~\ref{ppcorr1} (left panel) transformed to $p_t$ with the proper Jacobian.
} 
 \end{figure}

Figure~\ref{ptpt2} (left panel) shows the Glasma data in Fig.~\ref{ptpt1} (left panel) with the prefactor formulated using the gluon spectrum defined by Eq.~(\ref{glaseq}), which is much harder than the hadron spectrum. The result is still generally consistent with soft-component  \ptpt or \ytyt correlations.

%%%%%%%%%%
%%%%%%%%%%%%

Figure~\ref{ptpt2} (right panel) shows the \pp parametrization in Fig.~\ref{ppcorr1} (left panel) representing data in Refs.~\cite{porter2,porter3} replotted on $(p_{t1},p_{t2})$ with $p_t \in [0.2,5]$ GeV/c and with the proper Jacobian. The hard-component peak mode is near 1 GeV/c as expected, and the soft-component peak is just visible near the origin. The shape of the hard component  on \ptpt projected to 1D is in quantitative agreement with the hard component inferred directly from \pp single-particle $p_t$ spectra in Ref.~\cite{ppprd} (Fig.\ 10, left panel).

This comparison reveals that the Glasma flux-tube model is strongly contradicted by measured hadron \ptpt correlations. The Glasma histogram corresponds qualitatively to the soft component of measured hadron correlations. There is no corresponding hard-component peak, no large-angle parton scattering, in the Glasma model. The qualitative difference persists even when a prefactor derived from a hard gluon spectrum is introduced.

One might argue that the intervening hadronization process could invalidate such a comparison. The maximum hadron momentum per gluon would result from direct $1 \rightarrow 1$ correspondence as in local parton-hadron duality (LPHD)~\cite{lphd} (but observed local charge conservation is not then respected). Fragmentation ($1 \rightarrow 2$ or more) should actually reduce the mean hadron momentum relative to the Glasma spectrum. There is thus no possibility, within the Glasma model of ``bulk'' hadron production, to generate a counterpart to the observed \ptpt or \ytyt hard component.

%$\sqrt{\rho_{ref}(p_{t1},p_{t2})} = \exp(10 m_\pi)\times  \sqrt{\exp(-10 m_{t1}) \times \exp(-10 m_{t2})}$, where a thermal spectrum with slope parameter $T \approx 0.1$ GeV is assumed. 

%The gluon spectrum goes to $33 / p_t^4$ at large momentum. Remove factor 33 in $\sqrt{\rho_{ref}}$ below. Actually, factor $\Delta \eta$ to approximate $d^2 n_{ch} / d\eta$.

%Since we observe in Fig.\ 2 that the gluon per-pair amplitude is actually decreasing with increasing $y_t$ the corresponding result for the per-particle measure with additional single-particle spectrum factor would be negligible amplitudes everywhere except for  $p_t \ll 1$ GeV/c.

%$10^5 \exp(-10 m_{t1}) \times \exp(-10 m_{t2})$

%basically, $\kappa_2 \rightarrow$ constant after boost. So, $\Delta \rho / \sqrt{\rho_{ref}} \rightarrow \sqrt{\rho_{ref}}$, the soft component.

%$\kappa_2 = N_{part} \Delta \rho / \rho_{ref}$ so $\Delta \rho / \sqrt{\rho_{ref}} = [2 \sqrt{\rho_{ref}} / N_{part}] \kappa_2 / 2$ -- \pp only???

\subsection{Glasma  {$\bf p_t \times p_t$} correlations and radial boosts}

A proposed mechanism for formation of the SS 2D peak in angular correlations from more-central \auau collisions is radial boost of Glasma flux tubes. We can then ask what would be the effect of conjectured radial flow on the predicted $\kappa_2(p_{t1},p_{t2})$ in Fig.~\ref{ptpt1} (left panel)? Does radial flow also produce the nominal hard-component structure in \ptpt correlations, as in Fig.~\ref{ptpt2} (right panel)?

Starting with the 2D histogram in Fig.~\ref{ptpt1} (left panel) the following procedure was applied with two $p_t$ spectrum models $\rho_0(p_t)$ (prefactor $N_{FT}$ is ignored to simplify terminology, $\kappa_2 \rightarrow \Delta \rho / \rho_{ref}$). Reference $\rho_{ref}(p_{t1},p_{t2}) = \rho_0(p_{t1})\rho_0(p_{t2})$ was formed and numerator $\Delta \rho = \rho_{ref} \kappa_2$ was then obtained from the histogram in Fig.~\ref{ptpt1} (left panel). Azimuth angles were randomly sampled. $\Delta \rho$ and $\rho_{ref}$ were boosted from $(\vec p_{t1},\vec p_{t2})$ to $(\vec p'_{t1},\vec p'_{t2})$ by $\langle \beta_t\rangle = 0.6$ (maximum mean value inferred from central \auau collisions~\cite{radflow}). Boosted $\Delta \rho / \rho_{ref} \rightarrow \kappa_2$ was then recovered. The actual procedure was based on a Monte Carlo sampling ($10^7$ samples) of $\kappa_2$, $(p_{t1},p_{t2})$ and $(\phi_1,\phi_2)$ to construct boosted histograms. The two $\rho_0(p_t)$ spectrum shapes were a Maxwell-Boltzmann (M-B) with $T_{eff} = 0.15$ GeV and the gluon spectrum of Eq.~(\ref{glaseq}).

%%%%%%%%%%
 \begin{figure}[h]
  \includegraphics[width=1.65in,height=1.5in]{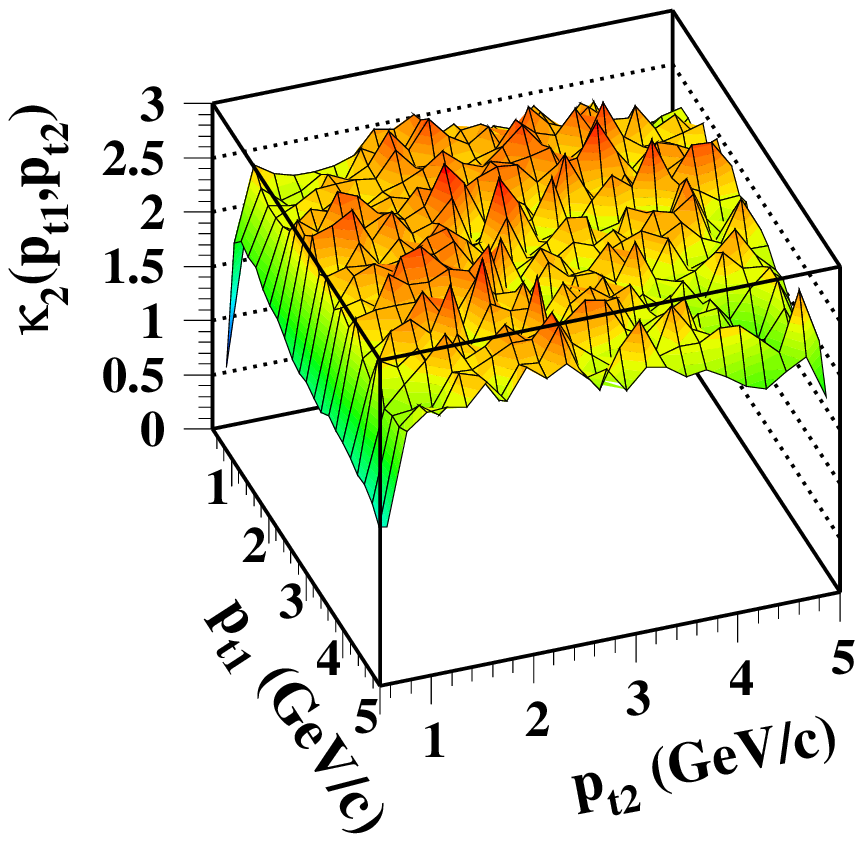}
  \includegraphics[width=1.65in,height=1.5in]{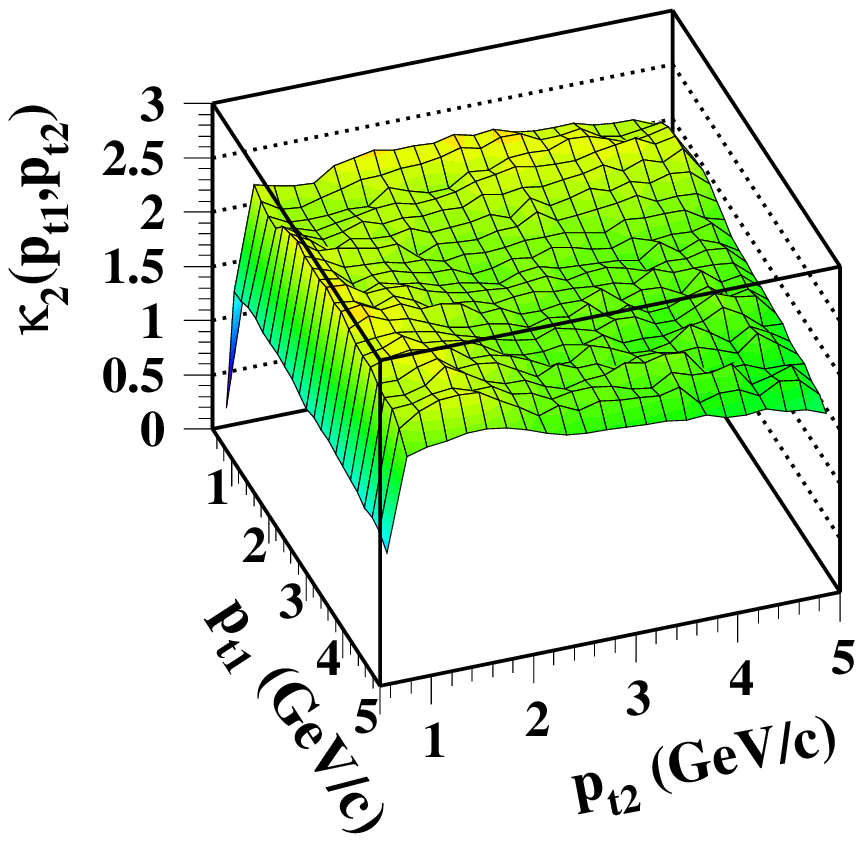}
\caption{\label{ytytfun} (Color online) Correlations from Fig.~\ref{ptpt1} (left panel) boosted by mean radial speed $\langle \beta_t \rangle = 0.6$ (see text).
Left: Single-particle spectrum is Maxwell-Boltzmann with $T_{eff} = 0.15$.
Right:  Single-particle spectrum is defined by Eq.~(\ref{glaseq}).
} 
 \end{figure}
%%%%%%%%%%%%

Results are shown in Fig.~\ref{ytytfun} for the two cases. For those boost conditions the $\kappa_2(p_{t1},p_{t2})$ distribution is changed modestly, mainly an increase at larger $p_t$ from 1 to 2 or 2.5 which cannot possibly match the hard-component structure evident in Fig.~\ref{ptpt2} (right panel) when converted to the per-particle form $\sqrt{\rho_{ref}(p_{t1},p_{t2})} \,\kappa_2(p_{t1},p_{t2})$. 

%It can be argued that the Glasma prediction is only relevant for more-central \auau collisions. However, the soft spectrum used to convert the histogram from Ref.~\cite{lappi} to the format of Fig.~\ref{ptpt2} has been boosted by rapidity shift $\Delta y_t = 0.6$ equivalent to the maximum radial flow inferred from central \auau collisions~\cite{radflow}. And $y_t \times y_t$ correlations for more-central \auau collisions are only quantitatively different from those for \pp collisions, with the expected factor 5 suppression at larger $y_t$ but much larger enhancement at smaller $y_t$ relative to \pp data~\cite{lanny}. Even for central \auau collisions measured $y_t \times y_t$ correlations strongly contradict the Glasma model because of the absence of transverse parton scattering and fragmentation in the model.

 \subsection{Glasma gluon azimuth correlations} \label{badcorr}

\begin{figure}[h]
  \includegraphics[width=1.65in,height=1.4in]{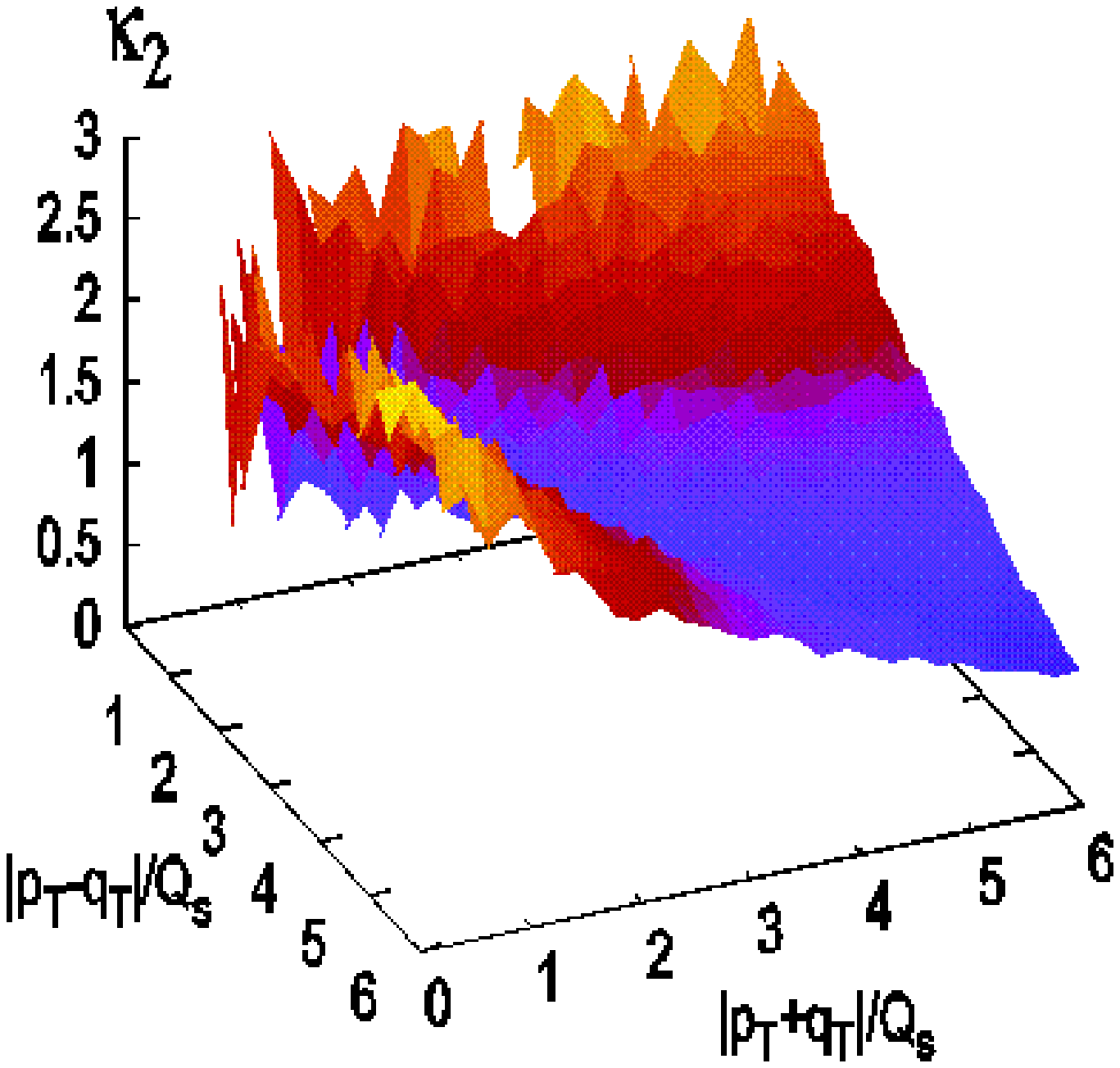}
  \includegraphics[width=1.65in,height=1.4in]{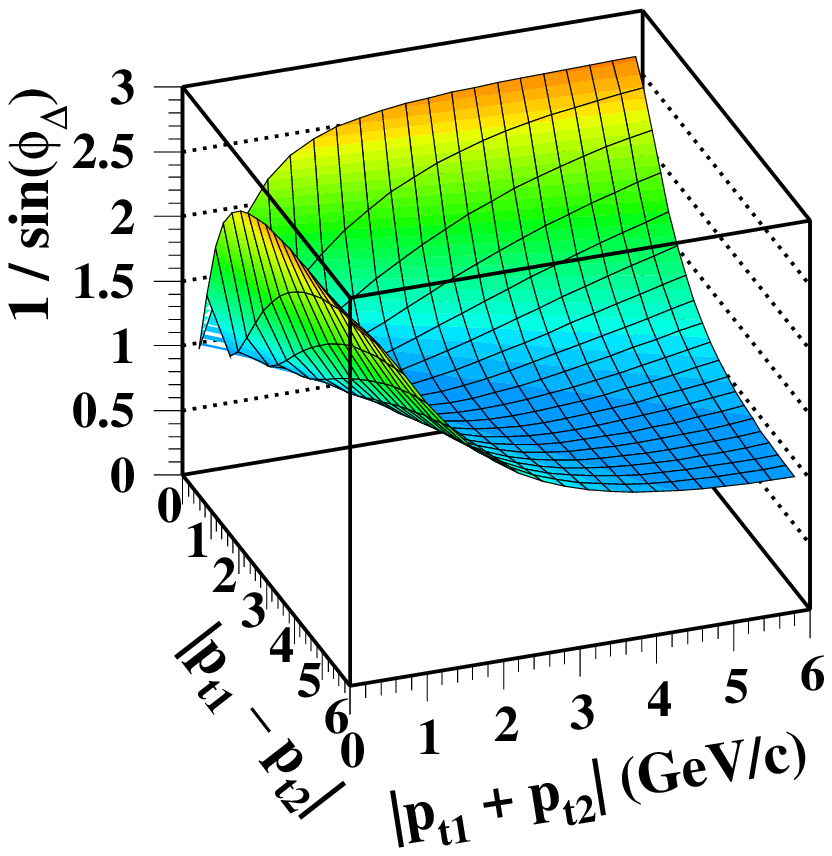}
\caption{\label{angcorr} (Color online)
Left: Histogram from Fig.~5 (right panel) of Ref.~\cite{lappi} showing two-gluon correlations on vector-momentum sum and difference.
Right: The Jacobian from $\phi_\Delta$ to $\cos(\phi_\Delta)$ plotted as a surface on $(x_+,x_-)$ (see text).
 } 
 \end{figure}

Figure~\ref{angcorr} (left panel) reproduces Fig.\ 5 (right panel) of Ref.~\cite{lappi} which is related indirectly to angular correlations plotted on $(\eta_\Delta,\phi_\Delta)$~\cite{porter2,porter3,axialci,daugherity}. The left panel in Ref.~\cite{lappi} includes extraneous self pairs (``large delta function peaks'') along one axis which are removed in the right panel. The horizontal axes are magnitudes of sum and difference vectors $\vec p_{t1} + \vec p_{t2}$ and $\vec p_{t1} - \vec p_{t2}$ relative to $Q_s = 1$ GeV.  One axis corresponds to parallel pairs, the other to antiparallel pairs.
Correlation structure seems to indicate a preference for momenta parallel and antiparallel (near the axes). 
The plot is interpreted to indicate that a ``collimation effect'' might be present in the initial state of \pp collisions, which might in turn explain the ``ridge'' observed in 7 TeV \pp collisions~\cite{cms}.

%%%%%%%%%%
%%%%%%%%%%%%

The coordinate axes in Fig.~\ref{angcorr} represented  by $x_+$, $x_-$ are defined by
\bea
x^2_\pm \equiv |\vec p_{t1} \pm \vec p_{t2}|^2 &=& 4 p_t^2 [1\pm \sin(2\psi)\cos(\phi_\Delta)]/2, %\\ \nonumber
%x_- = |\vec p_{t1} - \vec p_{t2}|^2 &\approx& 4 p^2 [1-\cos(\theta_{12})]/2
\eea
where $ p_{t1} = \sqrt{2}p_t \sin(\psi)$ and $ p_{t2} = \sqrt{2}p_t \cos(\psi)$ define $p_t$ and $\psi$ (polar coordinates in Fig.~\ref{ptpt1}).  The axes extend to $2p_{t,max} = 6$ GeV/c, twice the maximum single-particle momentum in the pair sample. The plotting variables can be inverted to $\sin(2\psi) \cos(\phi_\Delta) = (x^2_+ - x^2_-) / (x^2_+ + x^2_-) $ (normalized difference diagonal) and $4p_t^2 = x^2_+ + x^2_-$ (sum diagonal).
%, with $\sin(\theta_{12}) = 2x_+ x_- / (x^2_+ + x^2_-) $. 
%
The sum diagonal measures the quadratic mean of two transverse-momentum magnitudes. The difference diagonal measures angle $\phi_\Delta$ between pairs of momenta. For projection to space $(x_+,x_-)$ an average over $\sin(2\psi)$ would be determined by the (slowly varying) distribution in Fig.~\ref{ptpt1} (left panel). There is no sensitivity to elongation on $\eta_\Delta$, to a SS ridge per se. 

The apparent correlation structure is symmetric about $\phi_\Delta = \pi / 2$, and the angular distribution {\em seems to be}  sharply peaked toward the AS limit as well as toward the SS limit (``collimation effect even in the absence of radial flow''~\cite{lappi}), but the plot is misleading. Figure 9 (right panel) of Ref.~\cite{lappi2} shows the same correlation measure $\kappa_2$ plotted directly on $\phi_\Delta$ for $p_{t1} \approx p_{t2} \approx 3$ GeV/c [therefore $\sin(2\psi) \approx 1$]. The distribution on $\phi_\Delta$ is nearly uniform (variation within $\pm 10$\%). The relation between plots on $(x_+,x_-)$ and on $\phi_\Delta$ appears to be the Jacobian $1/\sin(\phi_\Delta)$ from $\phi_\Delta$ to $\cos(\phi_\Delta)$. 

Figure~\ref{angcorr} (right panel) shows the surface $1/\sin(\phi_\Delta)$ plotted on $(x_+,x_-)$.  The detailed agreement with the left panel is evident.
The structure in Fig.~\ref{angcorr} (left panel) therefore does not imply significant correlation on the actual azimuth difference $\phi_\Delta$, no inherent collimation effect in the initial state which might for example explain the SS ridge in \pp collisions at 7 TeV as suggested in Ref.~\cite{lappi}. In contrast, measured \pp hadron correlations depend strongly on $\phi_\Delta$ for hadron $p_t > 0.5$ GeV/c and weakly for $p_t < 0.5$ GeV/c (hard and soft components respectively), again contradicting the Glasma prediction.

\subsection{Glasma azimuth correlations and radial boosts}

A central issue for this study is the Glasma-model conjecture that the $\eta$-elongated SS 2D peak observed in minimum-bias angular correlations (``soft ridge'') represents formation of a 1D ridge on azimuth from boost of Glasma flux tubes emitting isotropically in their rest frames. Given the boost kinematics the peak on azimuth should be {\em narrower} for lower-$p_t$ gluons (hadrons), and conversely for higher-$p_t$ gluons (hadrons). That trend is opposite to what is actually observed for dihadron number correlations with applied $p_t$ cuts~\cite{porter2,ptcuts}, and for $p_t$ angular correlations~\cite{ptscale,ptedep}. 
In contrast, the observed data trends are well explained by jet formation, where higher-$p_t$ particles contribute a narrower SS peak structure. 
Boosted Glasma flux tubes are thus contradicted by $p_t$ systematics of the SS 2D peak width.
Furthermore, it was shown in Ref.~\cite{dumitru} that a very large and problematic radial boost velocity ($\beta_t = 0.96$) is required to achieve the narrow azimuth width observed in the data.

The detailed SS peak {\em shape} on azimuth also provides important evidence. The Glasma model requires averaging SS peak widths over a broad radial-boost distribution and a $p_t$ spectrum. Such averaging invariably leads to long tails relative to the fundamental peak distribution (e.g.\ Sec.~IV-A of Ref.~\cite{tzyam}). No such tails are observed in the data. A narrow 1D peak on azimuth with near-ideal Gaussian shape is observed for all \auau centralities~\cite{daugherity}.

 \subsection{Glasma gluon pseudorapidity correlations} 

In Fig.\ 4 (left panel) of Ref.~\cite{lappi} a Glasma prediction for gluon correlations on \deta is compared with measured \auau hadron correlations~\cite{phobos}. The Glasma prediction is essentially uniform on \deta for $\eta_\Delta \in [-4,4]$. A ``short-range'' correlation peak introduced from PYTHIA (\pp collisions) to accommodate the heavy ion data is extraneous to the Glasma prediction. No explanation is given for how a Glasma-based prediction of gluon correlations is related quantitatively to triggered dihadron correlations. Predicted variations at larger $\eta_\Delta$ are outside the $\eta$ acceptance of most detectors. 

Measured correlation structure on \deta provides essential model tests. 
For all minimum-bias ($p_t$-integral) angular correlations from all \auau centralities measured within the STAR TPC acceptance the SS 2D peak is consistent with a 2D Gaussian with large curvature on $\eta_\Delta$ (e.g.~\cite{axialci,daugherity}). In the static Glasma flux-tube scenario there is no variation on $\eta$. 
The measured SS curvature on \deta must then (in the Glasma model) result from strong $\eta$ dependence of conjectured radial flow which is said to drive appearance of the SS 2D peak. However, the nonjet quadrupole nominally associated with elliptic flow shows no such $\eta$ dependence within the STAR TPC acceptance~\cite{davidhq}. If a hydro interpretation is imposed on both the SS 2D peak (radial flow) and the nonjet quadrupole (elliptic flow) a major discrepancy emerges between two hydro manifestations. The Glasma model and jet-related \deta structure are further discussed in Ref.~\cite{tomglasma}.

\subsection{Glasma theory comparison with hadron data}

In Eq.~5.3 of Ref.~\cite{lappi2} angular correlation data in the form \drho~\cite{axialci,daugherity} are compared with Glasma flux-tube predictions in the form %$\Delta \rho / \rho_{ref} = \kappa_2 / N_{FT}$
\bea \label{glascompare}
\frac{\Delta \rho}{\sqrt{\rho_{ref}}}(\phi_\Delta = 0) &=& \frac{dN}{dy} \frac{\Delta \rho}{ \rho_{ref}}\left(\gamma_B -  \frac{1}{\gamma_B}\right) \\ \nonumber
&=& \frac{\kappa_2}{13.5\,\alpha_s}\left(\gamma_B -  \frac{1}{\gamma_B}\right),
\eea
where the LHS is evaluated as $1/\sqrt{2\pi \sigma^2_{\phi_\Delta}} = 0.62$ with $\sigma_{\phi_\Delta} = 0.64$. But that is the amplitude for a unit-normal Gaussian, not what was actually measured in Refs.~\cite{axialci,daugherity}. 
The boost factor including $\gamma_B$ is applied without justification to correlation amplitude $\Delta \rho/ \rho_{ref}$ which has no structure on $\phi_\Delta$. A SS 2D peak with large curvature on \deta is implicitly compared to a 1D ridge uniform on $\eta_\Delta$. And the Glasma prediction scales as $N_{part}$ whereas the measured peak amplitude increases faster than $N_{bin}$.

Since $\Delta \rho / \rho_{ref} = \kappa_2 / N_{FT}$, Eq.~(\ref{glascompare}) as written implies that the gluon density per flux tube is $(1/N_{FT})\,dN_g / dy = 1/13.5 \alpha_s  \approx 1/7$, since it is assumed that $\alpha_s \approx 0.5$ ($Q = 0.8$ GeV~\cite{ffprd}). However, if factor $1/2\pi$ which belongs in the first line (to match the LHS definition in Sec.~\ref{corrmeas}) is restored we obtain  $(1/N_{FT})\, dN_g / dy \approx 1$ compared to hadron $dn_{ch}/d\eta = 2.5$ for NSD \pp collisions (Lund string fragmentation).
If we now insert those values into the first line of Eq.~(\ref{glascompare}) we obtain
\bea 
\frac{\Delta \rho}{\sqrt{\rho_{ref}}}(\text {SS peak}) &=& \frac{1}{N_{FT}} \frac{dN_g}{dy} \frac{\kappa_2}{2\pi} \left(\gamma_B -  \frac{1}{\gamma_B}\right) \\ \nonumber
\text{or} ~~~\kappa_2 &\approx& \frac{0.7 \times 2 \pi}{\gamma_B -  \frac{1}{\gamma_B}}.
\eea
The $\kappa_2$ estimate is larger by factor $2\pi$ than that from Ref.~\cite{lappi2}. Aside from the comparison method and the assumed radial flow boost the Glasma prediction is a factor $2\pi$ too small relative to measured angular correlations, whatever the \auau centrality (unspecified in Ref.~\cite{lappi2}).

%%%%%%%%%%%
\section{Discussion}

According to Ref.~\cite{lappi} correlations best reveal the \aa initial state (IS)  dynamics (compared to integral yields and spectra). The Glasma picture of the IS in heavy ion collisions should be the natural framework to understand correlation mechanisms. In particular, Glasma flux tubes should explain the ``ridge'' encountered in more-central \auau collisions. To test that conjecture we contrast Glasma theory predictions of gluon correlations from Refs.~\cite{lappi,lappi2} with alternative models and with hadron fluctuation and correlation data.

%Constant theme is SS 2D peak vs Glasma FT and radial flow. Take that from Abstract and intro to Summary.

%Reference \cite{lappi} does provide valuable predictions because the Glasma results are so different from actual correlation measurements. The paper also reflects a degree of ignorance about progress in statistical and correlation analysis methods and  phenomenology established within the STAR collaboration throughout the RHIC program. 

%What is the meaning of nonperturbative strongly-interacting system with weak coupling constant ($\alpha_s \ll $?).

\subsection{Glasma and ``bulk'' hadron production}

%Several mechanisms have been proposed for hadron production at mid-rapidity in high-energy nuclear collisions. Is the dominant mechanism Glasma flux tubes, Lund strings, diffractive dissociation, large-angle gluon scattering, ``freezeout'' of a flowing partonic medium?  
%We consider a combination of correlation and spectrum data to reduce possibilities.

The CGC model describes ``bulk'' hadron production in terms of a longitudinal color-field system (Glasma flux tubes) approximating a dense gluonic system near mid-rapidity in \aa collisions. 
%Partons at larger $x$ form a color-charge source system for a longitudinal classical color field at smaller $x$---the glasma field. 
There is no transverse parton dynamics in the model, no large-angle parton scattering and fragmentation to jets. The Glasma model is a one-component (soft) model.
Mid-rapidity gluons are produced by emission from independent Bose-Einstein radiators (flux tubes) longitudinally boost invariant over some interval and with transverse correlation length $\sim 1 / Q_s$. 

The Glasma model is formally similar to the Lund string model~\cite{lund}. The probability distribution of large-$x$ color-charge field sources is analogous to a parton distribution function (PDF)~\cite{lappi}. However, the Glasma model does not provide an absolute prediction for hadron production, only relative trends on \aa centrality and energy. The relative centrality trend (Sec.~\ref{glascentral}) is contradicted by spectrum data~\cite{tomglasma}.

%There are $K \approx Q_s^2 S_\perp$ flux tubes, each flux tube radiating $n \sim 1/\alpha_s$ gluons. The gluon multiplicity is then $dN_g / d\eta \approx n K$. The charged-hadron multiplicity is similar (factor 2/3) according to local partin-hadron duality (LPHD) CITE.

%The glasma flux-tube number can be expressed in terms of \aa Glauber geometry parameters by $K \sim N_{part}$, proportional to the number of large-$x$ color charges CITE. 

In contrast, the Lund string model does provide quantitative predictions of  several aspects of nonperturbative soft particle production in \pp and more-peripheral \aa collisions~\cite{pythia,hijing}. The soft component observed in NSD \pp collisions~\cite{ppprd} (where ``bulk'' particle production is unlikely) appears to play a role even in central \auau collisions~\cite{hardspec}, describing spectrum data quantitatively when supplemented by a hard component (jet fragments) described by pQCD~\cite{fragevo}. In more-central \auau collisions the hard component contributes about one third of the total hadron production~\cite{jetspec}. Correlations in all collision systems require a two-component model including both soft production (Lund strings $\approx$ flux tubes) and (semi)hard parton scattering and fragmentation~\cite{porter2,daugherity}.

\subsection{Gluon  correlations inferred from hadron data}

Measured hadron correlations near mid-rapidity (Sec.~\ref{ppcorr}) exhibit apparent local charge and momentum conservation consistent with single-gluon parents common to two or more daughter hadrons, favoring a ($1 \rightarrow 2$) process. 
Hadron correlations seem to ``point back'' to single-parton (gluon) precursors. 
Hadron correlations from ($2 \rightarrow 1$) coalescence would require that several parent partons conspire to produce observed two-hadron local net-charge and momentum correlations.

Hadron spectrum structure can be combined with angular and \ptpt correlations consistent with the two-component model of hadron production 
 to reconstruct the parent parton population near mid-rapidity. 
%SOFT
Hadron correlation data suggest that partons from dissociated projectile nucleons comprise a minimally-correlated low-$p_t$ gluon population fragmenting {\em longitudinally} to charge-neutral hadron pairs which locally also conserve transverse momentum. That population produces the hadron correlation soft component which indicates no significant correlation among parent gluons. %The spectrum soft component in \aa collisions is consistent with that in \pp collisions as described by PYTHIA~\cite{porter2}.

%JETS

A scattered-gluon spectrum near mid-rapidity predicted quantitatively by pQCD~\cite{fragevo} can be combined with measured fragmentation functions~\cite{ffprd} to predict hadron fragment distributions consistent with measured hadron hard components~\cite{hardspec}, thus confirming a parton spectrum with lower bound near 3 GeV. Hadron correlation structure on \ptpt or \ytyt is consistent with that underlying parton spectrum.
Analysis of hadron angular correlations indicates that scattered partons are correlated as momentum-conserving back-to-back recoil pairs, including an acoplanarity distribution depending on the initial-state parton $k_t$ spectrum, but are otherwise uncorrelated. Even in central \auau collisions the $p_t$ spectrum of SS 2D peak hadrons and correspondence with the AS 1D ridge compel interpreting the SS 2D peak in terms of large-angle scattering of energetic ($> 3$ GeV) gluons.

%For partons (gluons) above some energy scale hadronization is described by parton (mainly gluon) fragmentation functions (FFs)~\cite{ffprd}.  

%make clear what are the hadron correlations from which gluon correlations are inferred.

%but the hadronization process may distort any true gluon correlations and may generate hadron correlations that do not correspond to true gluon correlations. Different results should be expected if hadronization proceeds via fragmentation ($1 \rightarrow n$), by coalescence ($n \rightarrow 1$) or according to local parton-hadron duality ($1 \rightarrow 1$).

\subsection{Glasma gluon correlations on $\bf p_t$ or $\bf y_t$}

%We now consider gluon correlations predicted by the Glasma flux tube model vs measured hadron correlations. 

Hadron correlations on \ptpt or \ytyt impose key constraints on any theoretical attempt to describe the SS 2D peak structure. $y_t \times y_t$ correlations from \pp collisions have a detailed quantitative correspondence to pQCD calculations and to spectrum hard components (Sec.~\ref{ppcorr}). Decomposition of $y_t \times y_t$ correlation structure according to  combinations of SS and AS azimuth subregions and LS and US charge combinations reveals quantitative correspondence with expected features of parton scattering and fragmentation~\cite{porter2,porter3,ffprd,ppprd,hardspec,fragevo}.

{\em Per-pair} measure $\kappa_2$ from Ref.~\cite{lappi} plotted in Fig.~\ref{ptpt1} already reveals major disagreement between Glasma gluons and hadron per-pair \pp correlations from Ref.~\cite{porter0}. Whereas the latter greatly increase in amplitude with larger $y_t$ the Glasma prediction in Fig.~\ref{ptpt1} generally decreases to a constant asymptotic value near unity. If both results are converted to per-particle measures as in Fig.~\ref{ptpt2} the large discrepancy is again apparent, but the measured hard-component structure appearing in the right panel~\cite{porter2,porter3} corresponds quantitatively with $p_t$ spectrum structure~\cite{ppprd,hardspec} and pQCD calculations~\cite{fragevo}.

Because it lacks a parton large-angle scattering mechanism the Glasma model cannot describe measured \ytyt correlations in \pp or more-central \auau collisions. The \auau case is quantitatively different from \pp collisions but still qualitatively incompatible with the Glasma model. Ironically, the Glasma model produces a very hard gluon $p_t$ spectrum as in Eq.~(\ref{glaseq}), but nothing that corresponds to the observed hadron spectrum hard component which scales with $N_{bin}$, not $N_{part}$.

\subsection{Glasma gluon angular correlations}

%Predicted gluon correlations are flat on $(\eta_\Delta,\phi_\Delta)$ in a static system. 

Figure~\ref{angcorr} (left panel) gives the impression that static-model gluon correlations 
%on $(\eta_\Delta,\phi_\Delta)$ 
are sharply peaked near parallel and antiparallel momentum configurations, suggesting that a SS ridge (``collimation'') is already inherent in the Glasma model without invoking radial flow.  
%But in other results from Glasma theory predicted correlation structure on $\phi_\Delta$ is negligible (less than 20\% variation), consistent with Fig.~9 (right panel) of Ref.~\cite{lappi2}.  The discrepancy is explained 
In Sec.~\ref{badcorr} the apparent peaked structure is identified as the Jacobian for the transformation $\phi_\Delta \rightarrow \cos(\phi_\Delta)$. 
Glasma correlations on \deta are also structureless within $|\eta_\Delta| < 4$.
%, per Fig.~4 of Ref.~\cite{lappi}. 
Thus, the static Glasma model predicts no significant gluon angular correlations for any conditions, in contrast to measured \pp and \auau hadron angular correlations.  

According to the theory any hadron correlation structure on azimuth must result from a conjectured radial boost. However, formation of a SS ridge via radial flow is inconsistent with hadron data. Coupling boost-invariant flux tubes with boost-invariant radial flow would produce correlations on $\phi_\Delta$ alone. The measured large SS curvatures on \deta would require strong variation of radial flow on $z$, contradicting the measured uniform nonjet azimuth quadrupole within the STAR TPC acceptance~\cite{davidhq}.  

The radial boost mechanism should produce a narrower structure on azimuth for lower-$p_t$ particles. The measured hadron SS peak is narrower on azimuth for {\em higher}-$p_t$ particles~\cite{porter2},  consistent with expectations for jet formation. There is no mechanism in the Glasma model for formation of the AS 1D peak on azimuth which tracks very closely with SS 2D peak properties vs \aa centrality. In contrast, the AS 1D peak and its systematics are consistent with expected back-to-back jet correlations.

\subsection{Glasma centrality dependence} \label{glascentral}

The centrality dependence of measured spectra and correlations from \auau collisions has been extensively studied~\cite{axialci,daugherity,lanny,hardspec,fragevo}.  The systematics of three SS 2D peak parameters challenge the Glasma flux-tube model.
%: peak amplitude, peak width on $\phi_\Delta$ and peak width or curvature on $\eta_\Delta$. 
The SS 2D peak (hard component) amplitude scales as $N_{bin}$ in more-peripheral \aa collisions (as predicted for jets) and increases more rapidly than $N_{bin}$ in more-central \auau collisions above a sharp transition on centrality~\cite{daugherity}. 
The Glasma flux-tube number $N_{FT}$ scales with \aa centrality as $Q_s^2 S_\perp \approx N_{part}$.  Any Glasma flux-tube contribution to the {\em per-hadron} SS 2D peak should actually {\em decrease} with increasing \aa centrality as $ N_{part}/ 2 n_{ch} \propto 1/[1 + x(\nu - 1)]$. The measured SS peak amplitude increases at least as fast as $\nu /  [1 + x(\nu - 1)]$~\cite{daugherity}. Thus, centrality dependence of the SS 2D peak amplitude is incompatible with Glasma expectations.

In the Glasma model the observed SS peak structure narrow on azimuth must result from conjectured radial flow (Sec.~\ref{badcorr}). But the magnitude of the radial boost and its effect on flux tube emission must conspire to produce the same SS 2D peak azimuth width $\sigma_\phi \approx 0.65$ over a broad range of \auau centralities where the reported radial flow magnitude $\beta_t$ is changing from 0.25 to 0.6~\cite{radflow}. And the azimuth width must remain constant through and beyond the sharp transition where the amplitude and $\eta$ width of the SS 2D peak (which must also depend on radial flow) change rapidly~\cite{daugherity}.

The conflicting constraints on conjectured radial flow systematics are inconsistent with observed nonjet quadrupole $v_2\{2D\}$ systematics which show no correspondence to the sharp transition in SS 2D peak characteristics~\cite{davidhq}. The nonjet quadrupole is uniform on $\eta_\Delta$ within the STAR TPC acceptance. Any radial flow would have to be very nonuniform within the same acceptance to produce the SS 2D peak curvature on $\eta_\Delta$. If the two phenomena are hydro manifestations they seem to be incompatible. A detailed search for radial flow with two-component spectrum analysis was unsuccessful~\cite{hardspec}.

%The Glasma (saturation-scale) model nominally applies to a dense \aa system.  In constrast, we observe smooth quantitative evolution of the SS peak from \pp to more-central \auau.

\subsection{Glasma energy dependence}

Per-particle fluctuation measure $(\sigma^2_N - \bar N) / \bar N = \bar N / k$ represents an integral of angular correlations~\cite{inverse}. Thus, NBD parameter $k$ in the form $\bar N / k$ represents a correlation integral of differential per-particle number angular correlations, including the SS 2D peak nominally described by Glasma flux tubes. The predicted energy trend for parameter $k$ in the Glasma model is $k  \approx N_{FT} \propto \sqrt{s}^\lambda$ increasing monotonically with $\sqrt{s}$, implying that $1/k$ should {\em decrease} slightly with energy. The observed trend in Fig.~\ref{fig1} is $1/k \propto [\ln(\sqrt{s} / \text{13.5 GeV})]^{0.75}$, and the SS 2D peak amplitude itself is observed to increase with energy approximately as $\ln(\sqrt{s} / \text{13.5 GeV})$~\cite{daugherity,ptedep}, strongly contradicting Glasma expectations.

%The Glasma theory predictions can be constrasted with data. With $N_{FT} \approx N_{part}$ slowly varying with energy we obtain from Fig.~\ref{fig1} (left panel) $\kappa_2 = N_{FT} / k \sim 1/k \approx  0.2\{\ln(\sqrt{s} / \text{13.5 GeV})\}^{0.75}$, strongly increasing with energy.

%The scale (angular acceptance) dependence of $\bar N/k$ (or equivalent) can be inverted to recover the underlying angular correlations~\cite{inverse,ptscale,ptedep}.

 \subsection{Glasma gluon fluctuations and correlations} \label{fluctcorr}

The statistics of generic two-tiered particle production can be described as follows: If $N$ particles result from event-wise production of $K$ sources, each emitting $n$ particles (all are fluctuating random variables) then~\cite{tomglasma}
\bea \label{twotier}
\frac{\sigma^2_N - \bar N}{\bar N}  &=&  \frac{\sigma^2_n - \bar n }{\bar n} + \lambda \frac{\sigma^2_{n_1 n_2}}{\bar n} + \bar n\, \left[\frac{\sigma^2_K - \bar K}{\bar K}\right] + \bar n,
\eea
illustrating the additivity of per-particle (co)variance measures~\cite{inverse}. It is assumed that $K$-$n$ covariance is zero. That expression could describe gluon emission from Glasma flux tubes, hadron emission from Lund strings or fragmentation of large-angle-scattered partons. 
%For simplicity we take $\bar N \approx \bar K \bar n$. 
%
The terms on the RHS are interpreted in terms of jet production in Ref.~\cite{tomglasma} (Sec.~V-A).
The first term represents {\em intra}\,source correlations, the second {\em inter}\,source correlations, the third represents non-Poisson source number fluctuations and the last term represents the hierarchy process apart from any source or particle correlations.
%The third and fourth terms represent correlations uniform on ($\eta_\Delta,\phi_\Delta$), not resolved peaks.

In the Glasma model the following are equivalent: $K \rightarrow k \approx N_{FT}  \approx N_{part}$ the number of flux tubes, $n \rightarrow n_g$, the number of gluons per independent color source, and $N \rightarrow N_g \approx n_g N_{FT} $ the total radiated gluon number. Independent color sources ($\approx$ flux tubes) are assumed to be Poisson distributed with no intersource correlations, in which case the second and third terms on the RHS of Eq.~(\ref{twotier}) are zero. We then have from Eq.~(\ref{fluxtube})
\bea \label{glasmeq}
\kappa_2 &=& N_{FT} \frac{\Delta \rho}{\rho_{ref}} \rightarrow N_{FT} \frac{\sigma^2_{N_g} - \bar N_g}{\bar N_g^2} \\ \nonumber
  &=&  \frac{\sigma^2_{n_g} - \bar n_g }{\bar n^2_g}  + 1,
\eea
where the first term in the second line is the per-pair measure of all intrasource (per flux tube) correlations, explaining why $\kappa_2$ tends to unity at larger $p_t$: intrasource correlations go to zero in the perturbative limit.
%, and significant model correlations are measured by $\kappa_2 - 1$. 
In the limit where exactly one gluon proceeds from one flux tube (no statistical hierarchy) $\sigma^2_{n_g} = 0$, $\bar n_g = 1$ and $\kappa_2 \rightarrow 0$.

Equation~(\ref{glasmeq}) contrasts dramatically with hadron data. $N_{FT} \approx N_{part}$, $\kappa_2 = [N_{part} / \rho_0(b)] [\rho_0(b) / k]$ and both factors have been measured. The first factor (centrality dependence) decreases $\propto 1 / [1 + 0.1(\nu - 1)]$~\cite{jetspec}. The second factor (energy dependence) plotted in Fig.~\ref{fig1} (right panel) increases strongly with $\sqrt{s}$ and is identified with several contributions from jet correlations in~\cite{tomglasma}.

%Need a small summary here. $\kappa_2 \geq 1$.

%In the context of a NBD the LHS is $\bar N / k$.  

%NBD parameter $k$ is described by $\kappa_2 = N_{FT} /k = 1 + (\sigma^2_{n_g} - \bar n_g) /\bar n^2_g $ or $k  \approx N_{FT} \approx N_{part}$, confirming that the $k$ parameter must have a weak energy dependence in the Glasma model. Measured $k$ values in  Fig.~\ref{fig1} reveal a strong energy dependence consistent with expectations for parton scattering and fragmentation to minijets.

\subsection{Glasma flux tubes vs minijets in \pp and \auau}

The mechanism for $\eta$-elongation of the SS 2D peak in more-central \auau collisions~\cite{axialci,daugherity} is a major problem for QCD theory at RHIC. Does the elongated peak arise from nonperturbative modification of parton scattering and fragmentation in a large \aa system, or does a novel process based on Glasma flux tubes and conjectured radial flow manifest as an elongated ridge?
%argument for:
Survival of copious parton scattering and fragmentation (minijets) in more-central \auau collisions~\cite{fragevo,tzyam,nohydro,jetspec} contradicts claims for formation of a flowing partonic medium with small viscosity~\cite{perfect3,perfect4}. Thus, an alternative explanation for the SS 2D peak by a mechanism other than parton scattering and fragmentation is sought.

The SS 2D peak has been characterized as a ``soft ridge''~\cite{gmm}, and $\eta$ elongation is described as ``long-range'' correlations, confusing polar angle measure $\eta$ and longitudinal momentum $p_z$ represented by longitudinal rapidity $y_z$. A causal argument is then invoked that only a process occuring at early times can produce long-range rapidity correlations.
Glasma flux tubes are said to be inherently long-range (boost invariant) and established at early times. They are thus characterized as a ``natural'' explanation for the $\eta$-elongated SS 2D peak~\cite{lappi}.

However, while the Glasma model seems to describe qualitatively a few features of the SS 2D peak in more-central \auau collisions, there are substantial discrepancies between the flux-tube model and measured hadron spectrum and correlation systematics.  Major issues include: (non)existence of conjectured radial flow and inconsistency with observed azimuth quadrupole systematics, disagreement with the energy and centrality dependence of hadron spectra and correlations, disagreement with most features of measured hadron $p_t$ and angular correlations, especially $\eta$ dependence of the SS 2D peak, absence of the \ptpt hard-component structure identified with the $\eta$-elongated SS 2D peak and absence of an AS 1D ridge in the Glasma model. 

In contrast, the two-component soft+hard (=minijet) model of hadron spectra and correlations has been quantitatively applied in combination with pQCD predictions to data from \pp and \auau collisions with good success~\cite{ppprd,hardspec,porter2,fragevo,jetspec}. The model continues to describe the SS 2D peak volume (sum of fragment pairs) for all \auau centralities. Large-angle parton scattering does occur at early times in the collision, with possible large-$\eta$ consequence in more-central \aa collisions. The only manifestation not currently accommodated by the two-component (fragmentation) model is $\eta$ elongation of the SS peak in more-central \aa collisions. 
However, deviations from pQCD expectations (e.g.\ angle asymmetries about the jet axis) are not inconsistent with conventional jet production in elementary collisions, for instance \pp collisions~\cite{porter3} and three-jet events in \ee collisions~\cite{aleph}. 

%Finally, the timeline seems to be unworkable. Flux tubes are an initial-state phenomenon, and gluon emission from those colored sources should proceed rapidly. Radial flow however should require a significant parton equilibration process and subsequent flux tube response to radial pressure~\cite{tomglasma}. 

% azimuth dependence:

%%%%%%%%%
 \section{Summary}

Observed strong elongation on $\eta$ of the same-side 2D peak in minimum-bias angular correlations from \auau collisions has been attributed to Glasma flux tubes coupled with radial flow to form a narrow structure or ridge on azimuth. In the present study we have tested that conjecture by comparing Glasma predictions for particle production, spectra and correlations with conventional fragmentation models and with measurements. We find a number of contradictions between the Glasma model and spectrum and correlation data.

The \aa centrality dependence of the Glasma model is defined by flux-tube number $N_{FT}$, approximated by the number of nucleon participants $N_{part}$. The Glasma one-component model of gluon (hadron) production therefore has no correspondence to observed hard-component features in spectra and correlations which scale with centrality as the number of \nn binary collisions $N_{bin}$.  There is no relation to \pp and peripheral \aa spectrum and correlation systematics described quantitatively by pQCD. 

The same-side 2D peak in angular correlations is associated with a prominent peak on \ptpt or \ytyt correlations with quantitative correspondence to the hard component in measured $p_t$ spectra, all scaling as $N_{bin}$. The hard-component peak remains visible near $p_t = 1$ GeV/c from \pp to central \auau collisions. 
The Glasma model predicts structure similar to observed {\em soft component} correlations which are unrelated to the same-side 2D peak. 
No hard-component structure on \ptpt or \ytyt correlations is observed from the model, and the Glasma single-gluon spectrum scaling with $N_{part}$ has no correspondence in hadron $p_t$ spectra.

The static Glasma model exhibits no significant angular correlations, relying on coupling to conjectured radial flow to develop a peak structure on azimuth.  However, such a correlation mechanism would produce peak azimuth-width dependence on particle $p_t$ (e.g.\ trigger-associated cuts) opposite to that observed (which is consistent with pQCD jet structure). Radial flow would also have to exhibit strong variation on $z$ to produce the large observed same-side peak curvature on $\eta_\Delta$, which would be inconsistent with measured nonjet quadrupole systematics. And the Glasma model has no mechanism to account for the away-side 1D peak on azimuth naturally explained by back-to-back jet correlations scaling as $N_{bin}$.

The Glasma model does provide a qualitative conjecture as to a possible mechanism for same-side peak $\eta$ elongation. But Glasma predictions are contradicted by measured hadron spectra and correlations. In contrast, a two-component model of hadron production, including minimum-bias parton fragmentation within a pQCD framework, quantitatively describes most hadron data. 
%A mechanism for $\eta$ elongation of jet structure in \aa collisions may soon be forthcoming within a pQCD model.

%%%%%%%%%%%%%%%%%%
\begin{acknowledgements}
This work was supported in part by the Office of Science of the U.S. DOE under grants DE-FG03-97ER41020 (TAT) and DE-FG02-94ER40845 (RLR).
\end{acknowledgements}

%%%%%%%%%%%%%%%%%%%%%%%%%%%%

\end{document}